\documentstyle[12pt]{article}
\def\be{\begin{equation}}
\def\ee{\end{equation}}
\def\bc{\begin{center}}
\def\ec{\end{center}}
\def\bea{\begin{eqnarray}}
\def\eea{\end{eqnarray}}
\def\da{{\dot{\alpha}}}
\def\db{{\dot{\beta}}}
\def\ha{{\hat{\alpha}}}
\def\hb{{\hat{\beta}}}
\def\cm{{\cal M}}
\def\cg{{\cal G}}
\def\ov{\overline}
\def\gev{{\rm \; GeV}}
\def\tev{{\rm \; TeV}}
\def\str{\rm \; Str \;}
\def\mpl{M_{\rm P}}
\def\mun{M_{\rm U}}
\def\msu{M_{\rm SUSY}}
\def\simlt{\stackrel{<}{{}_\sim}}
\def\str{{\rm Str \,}}
\def\zzbar{(z, \overline{z})}
\newcommand{\nsect}{\setcounter{equation}{0}
\def\theequation{\thesection.\arabic{equation}}\section}

\catcode`@=11
\def\marginnote#1{}
\newcount\hour
\newcount\minute
\newtoks\amorpm
\hour=\time\divide\hour by60
\minute=\time{\multiply\hour by60 \global\advance\minute by-\hour}
\edef\standardtime{{\ifnum\hour<12 \global\amorpm={am}%
        \else\global\amorpm={pm}\advance\hour by-12 \fi
        \ifnum\hour=0 \hour=12 \fi
        \number\hour:\ifnum\minute<10 0\fi\number\minute\the\amorpm}}
\edef\militarytime{\number\hour:\ifnum\minute<10 0\fi\number\minute}
\def\draftlabel#1{{\@bsphack\if@filesw {\let\thepage\relax
   \xdef\@gtempa{\write\@auxout{\string
      \newlabel{#1}{{\@currentlabel}{\thepage}}}}}\@gtempa
   \if@nobreak \ifvmode\nobreak\fi\fi\fi\@esphack}
        \gdef\@eqnlabel{#1}}
\def\@eqnlabel{}
\def\@vacuum{}
\def\draftmarginnote#1{\marginpar{\raggedright\scriptsize\tt#1}}
\def\draft{\oddsidemargin 0.0truein
        \def\@oddfoot{\sl preliminary draft \hfil
        \rm\thepage\hfil\sl\today\quad\militarytime}
        \let\@evenfoot\@oddfoot \overfullrule 3pt
        \let\label=\draftlabel
        \let\marginnote=\draftmarginnote
   \def\@eqnnum{(\theequation)\rlap{\kern\marginparsep\tt\@eqnlabel}%
\global\let\@eqnlabel\@vacuum}  }
\catcode`@=12
\begin{document}
\begin{titlepage}
\vspace*{-1cm}
\phantom{bla}
\hfill{CERN-TH.7192/94}
\\
\phantom{bla}
\hfill{LPTENS-94/12}
\\
\phantom{bla}
\hfill{UCLA/94/TEP13}
\\
\phantom{bla}
\hfill{hep-th/9405188}
\vskip 2.0cm
\begin{center}
{\Large\bf Mass formulae and natural hierarchy
\\ in string effective supergravities}\footnote{Work
supported in part by the US DOE under contract
No.~ DE--AT03--88ER40384, Task E, by the US NSF under
grant No.~PHY89--04035, and by the European
Union under contract No.~CHRX-CT92-0004.}
\end{center}
\vskip 1.5cm
\begin{center}
{\large Sergio Ferrara}, {\large Costas Kounnas}\footnote{On leave
from the Laboratoire de Physique Th\'eorique, ENS, Paris, France.},
{\large Fabio
Zwirner}\footnote{On leave from INFN, Sezione di Padova, Padua,
Italy.}
\\
\vskip .3cm
Theory Division, CERN \\
CH-1211 Geneva 23, Switzerland
\end{center}
\vskip 1cm
\begin{abstract}
\noindent
We study some conditions for the hierarchy $m_{3/2} \ll \mpl$ to
occur naturally in a generic effective supergravity theory. Absence of
fine-tuning and perturbative calculability require that the effective
potential has a sliding gravitino mass and vanishing cosmological
constant, up to ${\cal O}(m_{3/2}^4)$ corrections.
In particular, cancellation of quadratically divergent contributions
to the one-loop effective potential should take place, including the
`hidden sector' of the theory. We show that these conditions can be
met in the effective supergravities derived from four-dimensional
superstrings, with supersymmetry broken either at the string
tree-level via compactification,
or by non-perturbative effects such as gaugino condensation. A
crucial role
is played by some approximate scaling symmetries, which are remnants
of discrete target-space dualities in the large moduli limit. We
derive
explicit formulae for the soft breaking terms arising from this class
of
`large hierarchy compatible' (LHC) supergravities.
\end{abstract}
\vfill{
CERN-TH.7192/94
\newline
\noindent
May 1994}
\end{titlepage}
\setcounter{footnote}{0}
\vskip2truecm
\nsect{Introduction}

If one tries to extend the validity of an effective field theory to
energy scales much higher than its characteristic mass scale, and
quantum corrections appear carrying positive powers of the cut-off
scale $\Lambda$, one is faced with a scale hierarchy problem. The
typical example is the gauge hierarchy problem [\ref{hierarchy}] of
the Standard Model (SM) of strong and electroweak interactions, seen
as a low-energy effective field theory. When the SM is extrapolated
to cut-off scales $\Lambda \gg 1 \tev$, there is no symmetry
protecting the mass of the elementary Higgs field, and therefore the
masses of the weak gauge bosons, from large quantum corrections
proportional to $\Lambda$. The most popular solution to the gauge
hierarchy problem of the SM is [\ref{les}] to extend the latter to a
model with global $N=1$ supersymmetry, effectively broken at a scale
$\msu \simlt 1 \tev$. These extensions of the SM, for instance
[\ref{mssm}] the Minimal Supersymmetric Standard Model (MSSM), can be
safely extrapolated up to cut-off scales much higher than the
electroweak
scale, such as the supersymmetric unification scale $\mun \sim
10^{16} \gev$, the string scale $M_S \sim 10^{17} \gev$, or the
Planck scale $\mpl \equiv G_N^{-1/2} / \sqrt{8 \pi} \simeq 2.4 \times
10^{18} \gev$.

In view of the following discussion, we would like to take a closer
look at the properties that guarantee the stability of the gauge
hierarchy against quantum corrections in the MSSM and its variants.
Using a momentum cut-off $\Lambda$, the one-loop effective potential
for a generic theory reads [\ref{cww}]
\be
\label{veff}
V_1 = V_0 + {1 \over 64 \pi^2} \str \cm^0 \cdot \Lambda^4 \log
{\Lambda^2
\over \mu^2} + {1 \over 32 \pi^2} \str \cm^2 \cdot \Lambda^2 + {1
\over 64 \pi^2} \str \cm^4 \log {\cm^2 \over \Lambda^2}  + \ldots \,
,
\ee
where the dots stand for $\Lambda$-independent contributions, $\mu$
is the scale parameter, and
\be
\label{straccia}
\str \cm^n \equiv \sum_i (-1)^{2J_i} (2 J_i + 1) m_i^n
\ee
is a sum over the $n$-th power of the various field-dependent mass
eigenvalues
$m_i$, with weights accounting for the number of degrees of freedom and
the statistics of particles of different spin $J_i$. In eq.~(\ref{veff}),
$V_0$ is the classical potential, which in the case of the SM (and
of the MSSM) should contain mass terms at most of
the order of the electroweak scale. The quantum correction to the
vacuum energy with the highest degree of ultraviolet divergence is
the $\Lambda^4$ term, whose coefficient $\str \cm^0$ is always
field-independent, and equal to the number of bosonic minus fermionic
degreees of freedom. Being field-independent, this term can affect
the discussion of the cosmological constant problem (when the theory
is coupled to gravity), but does not affect the discussion of the
gauge hierarchy problem. Anyway, this term is always absent in
supersymmetric theories, which possess equal numbers of bosonic and
fermionic degreees of freedom. The second most divergent term in
eq.~(\ref{veff}) is the quadratically divergent contribution,
proportional to $\str \cm^2$. In the SM, $\str \cm^2$ depends on the
Higgs field, and induces a quadratically divergent contribution to
the Higgs squared mass, the well-known source of the gauge hierarchy
problem. An early attempt to get rid of the quadratically divergent
one-loop contributions to the SM Higgs squared mass consisted
[\ref{veltman}] in imposing the mass relation $[(\partial^2 /
\partial \varphi^2) \str \cm^2 (\varphi) ]_{\varphi = v} = 0$;
neglecting the light fermion masses, this amounts to requiring $3 m_H^2 +
6 m_W^2 + 3 m_Z^2 - 12 m_t^2 = 0$. It is clear (for recent
discussions, see e.g. [\ref{einhorn}] and references therein) that
such a requirement is modified at
higher orders in perturbation theory, since it amounts to a
relation among the dimensionless couplings of the SM that is not
stable under the renormalization group. A more satisfactory solution
of the problem is provided by $N=1$ global supersymmetry. For
unbroken $N=1$ global supersymmetry, $\str \cm^n$ is identically
vanishing for any $n$, due to the fermion-boson degeneracy within
supersymmetric multiplets [\ref{zumino}]. The vanishing of
$\str \cm^2$ persists, as
a field identity, if global supersymmetry is spontaneously broken in
the absence of anomalous $U(1)$ factors [\ref{fgp}]. Indeed, to keep
the gauge hierarchy stable it is sufficient that supersymmetry
breaking does not reintroduce field-dependent quadratically
divergent contributions to the vacuum energy. This still allows for a
harmless, field-independent quadratically divergent contribution to
the effective potential, and is actually used to classify the
so-called soft supersymmetry-breaking terms~[\ref{gg}].
In the case of softly broken supersymmetry, the
$\Lambda^2$ term of eq.~(\ref{veff}) only contributes to the
cosmological constant. With a typical mass splitting $\msu$
within the MSSM supermultiplets, the logarithmic term in
eq.~(\ref{veff}) induces corrections to the Higgs mass terms (before
minimization), which are at most $O(\msu^2)$: the hierarchy is then
stable if $\msu \simlt 1 \tev$.

To go beyond the MSSM, one must move to a more fundamental theory
with spontaneous supersymmetry breaking. The only possible candidate
for such a theory is $N=1$ supergravity coupled to
gauge and matter fields [\ref{cfgvp}], where (in contrast with the
case of global supersymmetry) the spontaneous breaking of local
supersymmetry is not incompatible with vanishing vacuum energy. In
$N=1$ supergravity, the spin $2$ graviton has for superpartner the
spin $3/2$ gravitino, and the only consistent way of breaking
supersymmetry is {\em spontaneously}, via the super-Higgs
mechanism.
One is then bound to interpret the MSSM as an effective low-energy
theory derived from a spontaneously broken supergravity~[\ref{bfs}].
The scale of
soft supersymmetry breaking in the MSSM, $\msu$, is related (in a
model-dependent way) to the gravitino mass $m_{3/2}$, which sets the
scale of the spontaneous breaking of local supersymmetry. One might
naively think that, whatever mechanism breaks local supersymmetry and
generates the hierarchy $m_{3/2} \ll \mpl$, the condition $\msu \sim
m_{3/2} \simlt 1 \tev \ll \mpl$ remains sufficient to guarantee the
stability of such a hierarchy against quantum corrections. To explain
why this expectation is generically incorrect, and to motivate the
present work, we need first to review some general facts about
spontaneously broken $N=1$ supergravity.

Even barring higher-derivative terms, the general structure of $N=1$
supergravity still allows for a large amount of arbitrariness. First,
one is free to choose the field content. Besides the gravitational
supermultiplet, containing as physical degrees of freedom the
graviton and the gravitino, one has a number of vector
supermultiplets, whose physical degrees of freedom are the spin $1$
gauge bosons $A_\mu^a$ and the spin $1/2$ Majorana gauginos
$\lambda^a$, transforming in the adjoint representation of the chosen
gauge group. One is also free to choose the number of chiral
supermultiplets, whose physical degrees of freedom are spin $1/2$
Weyl fermions $\chi^I$ and complex spin $0$ scalars $z^I$, and their
transformation properties under the gauge group. Furthermore, one has
the freedom to choose a real gauge-invariant K\"ahler function
\be
\label{gfun}
\cg (z,\ov{z}) = K (z,\ov{z}) + \log |w(z)|^2 \, ,
\ee
where $K$ is the K\"ahler potential whose second derivatives
determine the kinetic terms for the fields in the chiral
supermultiplets, and $w$ is the (analytic) superpotential. One can
also choose a second (analytic) function $f_{ab} (z)$, transforming
as a symmetric product of adjoint representations of the gauge group,
which determines the kinetic terms for the fields in the vector
supermultiplets, and in particular the gauge coupling constants and
axionic couplings,
\be
g_{ab}^{-2} =  {\rm Re \,} f_{ab}  \, ,
\;\;\;\;\;
\theta_{ab} =  {\rm Im \,} f_{ab}  \, .
\ee
Once the functions $\cg$ and $f$ are given, the full supergravity
Lagrangian is specified. In particular (using here and in the
following the standard supergravity mass units in which $\mpl=1$),
the classical scalar potential reads
\be
\label{pot}
V = V_F + V_D = e^\cg \left( \cg^I \cg_I - 3 \right) + {[({\rm Re \,}
f)^{-1}]^{ab} \over 2} \left( \cg_I T^{I}_{a \;\;\; \ov{J}}
\ov{z}^{\ov{J}} \right) \left( z^L T^{\;\;\;\;\; \ov{K}}_{b \, L}
\cg_{\ov{K}} \right) \, .
\ee
In our notation, repeated indices are summed, unless otherwise stated;
we use Hermitian generators, $[(T_a)^I_{\;\;\; \ov{J}}]^\dagger =
T^{J}_{a \;\;\;\;\; \ov{I}}$; derivatives of the K\"ahler function
are denoted by $\partial \cg / \partial z^I \equiv \partial_I \cg
\equiv \cg_I$ and $\partial \cg / \partial \ov{z}^{\ov{I}} \equiv
\partial_{\ov{I}} \cg \equiv \cg_{\ov{I}}$; and the K\"ahler metric
is $\cg_{I \ov{J}} = \cg_{\ov{J} I} = K_{I \ov{J}} = K_{\ov{J} I}$.
The inverse K\"ahler metric $\cg^{I \ov{J}}$, such that $\cg^{I
\ov{J}} \cg_{\ov{J} K} = \delta^I_K$, can be used to define
\be
\cg^I \equiv \cg^{I \ov{J}} \cg_{\ov{J}}
\, ,
\;\;\;\;\;
\cg^{\ov{I}} \equiv \cg_{J}\cg^{J \ov{I}}
\, .
\ee
Notice that the $D$-term part of the scalar potential is always
positive semi-definite, $V_D \ge 0$, as in global supersymmetry.
However, in contrast with global supersymmetry, the $F$-term part of
the scalar potential is not positive semi-definite in general. On the
one hand, this allows for spontaneous supersymmetry breaking with
vanishing classical vacuum energy, as required by consistency with a
flat background. On the other hand, the requirement of vanishing
vacuum energy imposes a non-trivial constraint on the structure of
the theory,
\be
\label{vacuum}
\langle \cg^I  \cg_I \rangle = 3
\;\;\;\;\;
{\rm if}
\;\;\;\;\;
\langle V_D \rangle = 0  \, .
\ee

The order parameter of local supersymmetry breaking in flat space
is the gravitino mass,
\be
\label{mgrav}
m_{3/2}^2 \zzbar =  e^{\cg \zzbar} = |w(z)|^2 e^{K \zzbar}  \, ,
\ee
which depends on the vacuum expectation values of the scalar fields
of the theory, determined in turn by the condition of minimum vacuum
energy. The goldstino $\tilde{\eta}$ is given by
\be
\label{goldstino}
\tilde{\eta} =  e^{\cg \over 2} \cg_I \chi^I + {1 \over 2}
\cg_I T^{I}_{a \;\;\; \ov{J}} \ov{z}^{\ov{J}} \lambda^a \, .
\ee
On the right-hand side of eq.~(\ref{goldstino}), the two
contributions are associated with $F$-- and $D$--term breaking,
respectively. In the following, we shall assume that the $D$ breaking
is absent at tree level, as is the case in all interesting situations.
For convenience, we shall also classify the fields as $z^I \equiv (
z^\alpha,z^i)$, where the fields with Greek indices have
non-vanishing projections along the goldstino direction, $\langle
\cg^\alpha \cg_\alpha \rangle \ne 0$ (not summed), whereas the fields
with small Latin indices have vanishing goldstino projection,
$\langle \cg^i \cg_i \rangle = 0$. With these conventions,
eq.~(\ref{vacuum}) can be split as
\be
\langle \cg^\alpha  \cg_\alpha \rangle = 3 \, ,
\;\;\;\;\;
\langle \cg^i  \cg_i \rangle = 0 \, ,
\ee
and the goldstino just reads
\be
\label{gold}
\tilde{\eta} =  e^{\cg \over 2} \cg_I \chi^I
=  e^{\cg \over 2} \cg_{\alpha} \chi^{\alpha} \, .
\ee

If $N=1$ local supersymmetry is spontaneously broken on a flat
background\footnote{We recall that, on non-flat backgrounds,
stable vacuum states do not necessarily correspond to minima of the
vacuum energy [\ref{weinberg}]. For
the calculation of quadratic divergences on arbitrary backgrounds, see
ref.~[\ref{bc}].}, the coefficient of the one-loop quadratically divergent
contributions to the vacuum energy is given by [\ref{grk}]
\be
\label{genstr}
\str \cm^2 \zzbar = 2 \, Q \zzbar \, m_{3/2}^2 \zzbar
\, ,
\ee
where
\be
\label{qexpr}
\begin{array}{ccl}
Q \zzbar & = & N_{TOT} - 1 - \cg^I  \zzbar H_{I \ov{J}} \zzbar
\cg^{\ov{J}} \zzbar
\\
& = & N_{TOT} - 1 - \cg^\alpha  \zzbar H_{\alpha \ov{\beta}} \zzbar
\cg^{\ov{\beta}} \zzbar \, ,
\end{array}
\ee
\be
H_{I \ov{J}} \zzbar = R_{I \ov{J}} \zzbar + F_{I \ov{J}} \zzbar
\, ,
\ee
\be
\label{ricci}
R_{I \ov{J}} \zzbar \equiv \partial_{I} \partial_{\ov{J}}
\log \det \cg_{M \ov{N}} \zzbar \, ,
\ee
\be
\label{ficci}
F_{I \ov{J}} \zzbar \equiv - \partial_{I} \partial_{\ov{J}}
\log \det {\rm Re \,} [ f_{ab} (z) ] \, .
\ee
Clearly, the only non-vanishing contributions to $\str \cm^2$ come
from the field directions $z^\alpha$ for which $\langle \cg^\alpha
\cg_\alpha \rangle \ne 0$ (not summed). In eq.~(\ref{ricci}), $R_{I
\ov{J}}$ is the Ricci tensor of the K\"ahler manifold for the chiral
multiplets, whose total number is denoted by $N_{TOT}$. In
eq.~(\ref{ficci}), $F_{I \ov{J}}$ has also a geometrical
interpretation, since the way it is constructed from the gauge field
metric is very similar to the way $R_{I \ov{J}}$ is constructed from
the K\"ahler metric. It is important to observe that both $R_{I
\ov{J}}$ and $F_{I \ov{J}} $ {\em do not depend at all} on the
superpotential of the theory, but only depend on the
metrics for the chiral and gauge superfields. This very fact allows
for the
possibility that, for special geometrical properties of these two
metrics, the dimensionless quantity $Q \zzbar$ may turn out to be
field-independent and hopefully vanishing.

In a general spontaneously broken $N=1$ supergravity, the
non-vanishing of
$Q \zzbar$ induces, at the one-loop level, a contribution to the
vacuum energy
quadratic in the cut-off $\Lambda$. This leads to a very
uncomfortable situation, not only in relation with the cosmological
constant problem (a vacuum energy of order $m_{3/2}^2 \Lambda^2$
cannot be cancelled by any physics at lower energy scales)
but also in relation with the gauge hierarchy problem, which asks for
a gravitino mass not much larger than the electroweak scale. Since
$m_{3/2} \zzbar$ is a field-dependent object, and its expectation
value must arise from minimizing the vacuum energy, quadratically
divergent loop corrections to the latter may generically destabilize
[\ref{destab}] the desired hierarchy $m_{3/2} \ll \Lambda$,
attracting the gravitino mass either to $m_{3/2}=0$ (unbroken
supersymmetry) or to $m_{3/2} \sim \Lambda$ (no hierarchy).
This destabilization problem cannot be
solved just by moving from the cut-off-regulated supergravity to the
quantum supergravity defined by four-dimensional superstrings
[\ref{fds}], since the only practical difference\footnote{We shall
comment later on the possible contributions coming from string modes
that remain massive
in the limit of unbroken supersymmetry, and therefore do not appear in
the effective theory below the string scale.} will be to replace the
cut-off scale $\Lambda$ by an effective scale of order $M_S$. In a
generic supergravity theory, we
still have the freedom to evade this problem, by postulating the
existence of an extra sector of the theory, which gives an opposite
contribution to $Q$, so that $Q + \Delta Q = 0$. Such a request,
however, is very unnatural, and implies a severe fine-tuning among
the parameters of the old theory and of the extra sector. In
string-derived supergravities, the possibility of such a cheap way
out is lost, since all the degrees of freedom of the theory are known
and the total contribution to $Q$ is well defined. We no longer have
the freedom to compensate a non-zero $Q$ by modifying the theory!

{}From the previous discussion, it is clear that a satisfactory
solution of the hierarchy problem ($m_{3/2} \ll \mpl$), and the
perturbative stability of the flat background, at least up to
${\cal O} (m_{3/2}^4)$ corrections, require the vanishing of $Q
\zzbar$. It is also clear that, if such a solution
exists, this will put strong constraints on the
scalar and gauge metrics, see eqs.~(\ref{genstr})--(\ref{ficci}).
In order to appreciate the geometrical meaning of the vanishing
of $Q \zzbar$, we present here a simple working example (another,
string-motivated example was previously given in [\ref{fkpz1}]).
Consider a model containing $N_{TOT} \equiv N_c + 3$ chiral
superfields, three gauge singlets $(T,U,S)$ and $N_c$ charged
fields $C_i$ ($i=1,\ldots,N_c$), with a gauge kinetic function given
by $f_{ab} = \delta_{ab} S$, a K\"ahler function parametrizing a
$SU(1,N_c+1)/[U(1) \times SU(N_c+1)] \times SU(1,1) / U(1) \times
SU(1,1) / U(1)$ manifold,
\be
\cg =
- 3 \log (T + \ov{T} - C_i \ov{C}_i)
- k \log (U + \ov{U}) - \log (S + \ov{S})
+ \log | w(C,U,S) |^2 \, ,
\ee
and a superpotential $w(C,U,S)$, which depends non-trivially on all
fields apart from the singlet field $T$. One can easily prove that,
thanks to the field identity $\cg^T \cg_T \equiv K^T K_T \equiv 3$,
the scalar potential of such a model is automatically positive
semi-definite, with a flat direction along the $T$-field, as in the
`no-scale' models of refs.~[\ref{cfkn},\ref{ekn}]. At the minima
that preserve the gauge symmetry, $\cg_S = \cg_U = \cg_C = 0$,
whereas $\cg_T \ne 0$. The gauge coupling constant at the minimum is fixed
to the value $g^2 = ({\rm Re \, } S)^{-1}$, and the
VEV of the $U$ field is also fixed by the minimization condition,
whereas the gravitino mass $m_{3/2}^2= |w|^2 / [(S + \ov{S})
(T + \ov{T})^3 (U + \ov{U})^k]$ is classically undetermined, sliding
along the $T$ flat direction.  To compute $Q \zzbar$ in this model,
it is sufficient to realize that the Ricci tensors for the three
factor manifolds have the simple expressions ($I=0,1,\ldots,N_c$):
\be
R_{I \ov{J}} = {N_c+2 \over 3} \cg_{I \ov{J}} \, ,
\;\;\;\;\;
R_{S \ov{S}} = 2 \cg_{S \ov{S}} \, ,
 \;\;\;\;\;
R_{U \ov{U}} = {2 \over k} \cg_{U \ov{U}} \, .
\ee
By just applying eqs.~(\ref{qexpr})--(\ref{ficci}) to the present
case, we find
\be
Q = (N_c+3) - 1 - \cg^T R_{T \ov{T}} \cg^{\ov{T}} = N_c+2 - {N_c+2
\over 3}
\cg^T \cg_T = 0 \, .
\ee
In this simple example, we can clearly see that the vanishing of $Q
\zzbar$ occurs at all minima of the potential along the flat
direction $T$, and is completely independent of the details of the
superpotential.

In the present work, we discuss how the previously mentioned
conditions for the hierarchies $m_{3/2} \ll \mpl$ and $\langle
V \rangle \simlt {\cal O} (m_{3/2}^4)$ can be
realized in a generic effective supergravity theory. In particular,
we go beyond the example of ref.~[\ref{fkpz1}], and identify a wide
class of `large hierarchy compatible'
(LHC) models where, modulo ${\cal O}(m_{3/2}^2/\mpl^2)$ corrections,
\be
\label{hc}
\cg^{\alpha} \cg_{\alpha} = 3
\;\;\;\;\;
{\rm \bf and}
\;\;\;\;\;
Q = 0 \, .
\ee
In section 2, we discuss the problem at the pure supergravity level,
showing that the conditions of eq.~(\ref{hc}) can be naturally
achieved, as field identities, whenever the metrics for the gauge and
chiral superfields are compatible with some (approximate) scaling
properties. We also review some general mass formulae of
supergravity, and explore the specific form they take in the case of
the LHC models. In particular, we give explicit formulae for the
resulting mass parameters at the level of the MSSM, which are
subject to important restrictions and exhibit remarkable universality
properties\footnote{Results partially overlapping with ours were
obtained in [\ref{savoy}--\ref{bim}].}. In section~3, we first
review how the desired scaling
properties naturally appear in the effective supergravity theories,
which are extracted [\ref{effcl},\ref{efferm}] from four-dimensional
superstring constructions in the low-energy limit, under the
assumption that the fields $z^\alpha$, which trigger supersymmetry
breaking, have large VEVs compared to the string scale. This is
probably necessary in
order to obtain a gravitino mass much smaller than the string scale,
and is equivalent to neglecting the effects of winding modes in the
low-energy effective Lagrangian. In the same limit, one recovers an
approximate Peccei-Quinn symmetry for the moduli dependence of the
overall scalar field metric $\cg_{I \ov{J}}$ of any chiral multiplet.
This is consistent with the fact that this symmetry is restored when
non-trivial topological effects on the world-sheet (exponentially
suppressed [\ref{it}] in the large-volume limit for the internal
space) can be neglected. We then apply the formulae of section~2 to
a number of examples,
corresponding to the effective supergravity theories of different
four-dimensional superstring models (Calabi-Yau, orbifolds,
fermionic constructions, \ldots), and to different mechanisms for
supersymmetry breaking (coordinate-dependent compactifications
[\ref{ss1}--\ref{ss5}], gaugino condensation [\ref{gcond1}--\ref{lust}],
\ldots). In the final section, we
critically rediscuss the interpretation of our results, in particular
the role of string massive modes and higher-loop contributions to the
effective potential, and describe some prospects for further work.

\nsect{Supergravity mass formulae}

\subsection{General case}

We assume here, as announced in the Introduction, spontaneous
breaking of local $N=1$ supersymmetry with vanishing vacuum energy
and unbroken gauge symmetries (singlet goldstino). Before
specializing
to the case of LHC models, we would like to recall some general
supergravity formulae for the boson and fermion mass matrices.

The expression for the gravitino mass, $m_{3/2}$, has already been
given in eq.~(\ref{mgrav}). The mass matrices for the spin $1/2$
fermions are
\be
\label{mgaug}
(M_{1/2})_{ab} = { \cg^K f_{ab,K} \over 2} m_{3/2}
= { \cg^{\alpha} f_{ab,\alpha} \over 2} m_{3/2}
\ee
for the gauginos, and, after projecting out the goldstino-gravitino
mixing term, associated with the super-Higgs mechanism,
\be
\label{mchir}
(M_{1/2})_{IJ} =  \left( \cg_{IJ} -  \cg_{IJ\ov{K}} \cg^{\ov{K}}
+ {1 \over 3} \cg_I \cg_J  \right) m_{3/2} \equiv  \left( D_I \cg_J +
{1
\over 3} \cg_I \cg_J  \right) m_{3/2}
\ee
for the fermions in the chiral supermultiplets, where the covariant
derivative $D_I$ is defined with respect to the K\"ahler connection
$\cg^M_{JK} \equiv \cg^{M \ov{L}} \cg_{JK\ov{L}}$. The spin 0 mass
matrices are
\be
\label{anaanti}
(M_0^2)_{I \ov{J}} \equiv D_I D_{\ov{J}} V  = V_{I \ov{J}}
\ee
and
\be
\label{anaana}
(M_0^2)_{IJ} \equiv D_I D_J V =  V_{IJ} - \cg^K_{IJ} V_K \, .
\ee
In eqs.~(\ref{mgaug})--(\ref{anaana}), we have written the different
mass matrices as they appear in the supergravity Lagrangian: to
compute the physical mass eigenvalues, one should not forget the
presence of non-canonical kinetic terms, and rescale these
expressions by the appropriate powers of the gauge and K\"ahler
metrics, $[({\rm Re \,} f)^{-1/2}]^{cd}$ and $(\cg^{-1/2})^{K
\ov{L}}$
for each gauge and chiral index, respectively. In the following,
it will be useful to consider the fermion squared mass matrices,
\be
\label{mgsq}
\left( M_{1/2} M^{\dagger}_{1/2} \right)_{ab} \equiv (M_{1/2})_{ac}
[({\rm Re \,} f)^{-1}]^{cd} (M^{\dagger}_{1/2})_{db}   \, ,
\ee
\be
\label{mfsq}
\left( M_{1/2} M^{\dagger}_{1/2} \right)_{I\ov{J}} \equiv
(M_{1/2})_{IK}
\cg^{K \ov{L}} (M^{\dagger}_{1/2})_{\ov{L}\ov{J}}   \, .
\ee
Remembering the general expression of the scalar potential,
eq.~(\ref{pot}),
and the condition for a minimum with vanishing vacuum energy,
\be
\label{condi}
V =  V_I = 0 \, ,
\ee
we can reexpress the spin 0 mass matrices as
\be
(M_0^2)_{I \ov{J}} = \left( M_{1/2} M^{\dagger}_{1/2}
\right)_{I\ov{J}}
+ \left( \cg_{I \ov{J}} - \cg_K R^K_{\ov{J} I L} \cg^L + {1 \over 3}
\cg_I \cg_{\ov{J}} \right) m_{3/2}^2 \, ,
\ee
and
\be
(M_0^2)_{IJ} = V_{IJ} = \left[ ( \cg^K D_K + 2 ) (\hat{M}_{1/2})_{IJ}
\right] m_{3/2}^2 \, ,
\ee
where $R^K_{\ov{J}IL} = \partial_{\ov{J}} \cg_{IL}^K$ and
$(\hat{M}_{1/2})_{IJ}
= (M_{1/2})_{IJ} / m_{3/2}$. Incidentally, we can notice that
eq.~(\ref{condi}) implies
\be
\cg^I (M_{1/2})_{IJ} \, m_{3/2} = V_L - \cg_L V = 0 \, ,
\ee
consistently with the fact that the fermion mass matrix must have a
vanishing
eigenvalue with eigenvector in the goldstino direction.

Before concluding this section, we would like to give more explicit
formulae for the mass terms in the sector of the theory corresponding
to the fields $z^{i}$, for which  $\langle \cg^{i} \rangle = \langle
\cg_{i} \rangle = 0$. In realistic supergravity models satisfying
our assumptions, such a sector should include the chiral superfields of
the MSSM. One can easily find
\be
(M_{1/2})_{ij} = \left( \cg_{ij} -  \cg_{ij\ov{\alpha}}
\cg^{\ov{\alpha}} \right) m_{3/2} \, ,
\ee
\be
(M_0^2)_{i \ov{j}} = \left( M_{1/2} M^{\dagger}_{1/2} \right)_{i
\ov{j}}
+ \left( \cg_{i \ov{j}} - \cg_{\alpha} R^{\alpha}_{\ov{j} i
\beta} \cg^{\beta} \right) m_{3/2}^2 \, ,
\ee
\be
(M_0^2)_{i j} = V_{i j} = \left[ ( \cg^{\alpha} D_{\alpha} + 2 )
(\hat{M}_{1/2})_{i j} \right] m_{3/2}^2  \, .
\ee
In the sector under consideration, one can obtain a particularly
simple expression also for
\be
D_{i} D_{j} D_k V = V_{i j k} =  \left[ ( \cg^{\alpha} D_{\alpha}
+ 3 ) \hat{V}_{i j k} \right] m_{3/2}^2  \, ,
\ee
where
\be
\hat{V}_{i j k} \equiv D_{i} D_{j} D_k  (\cg^L \cg_L) = w_{i j
k} \, .
\ee

\subsection{LHC models}

We begin this section by considering a restricted theory with only
the $n$ fields $z^\alpha$, and a K\"ahler function of the form
\be
G (r^{\alpha}) = - \log Y (r^{\alpha}) + \log |w|^2 \, ,
\ee
where $Y$ is a homogeneous function of degree $p$, depending only on
the combinations
\be
\label{reals}
r^\alpha \equiv z^\alpha + \ov{z}^{\ov{\alpha}} =  2 {\rm Re \,}
z^\alpha
\, ,
\ee
and $w$ is assumed not to depend on $z^{\alpha}$, $\partial w /
(\partial z^{\alpha}) = 0$. In other words, we shall assume from now
on that
\be
\label{homo}
r^\alpha Y_\alpha = p \, Y  \, ,
\ee
where it is unambiguous to define $Y_\alpha \equiv \partial Y /
(\partial r^{\alpha}) \equiv \partial Y / (\partial z^{\alpha})
\equiv \partial Y / (\partial \ov{z}^{\ov{\alpha}})$. From
eq.~(\ref{homo}), it immediately follows that the K\"ahler metric
for the fields $z^{\alpha}$ is a homogeneous function of degree
$(-2)$,
\be
\label{zero}
r^{\gamma} G_{\alpha \beta \gamma} = - 2 \, G_{\alpha \beta} \, ,
\ee
and that
\be
\label{uno}
G^{\alpha} = - r^{\alpha} \, ,
\ee
\be
\label{due}
G^{\alpha} G_{\alpha} = p \, ,
\ee
\be
\label{tre}
G^{\alpha} R_{\alpha \beta} G^{\beta} = 2 n \, .
\ee
For $p=3$, one obtains the structure of the `no-scale' models:
for a non-vanishing $w$, supersymmetry is broken with
vanishing tree-level vacuum energy, and the gravitino mass,
$m_{3/2}^2 = |w|^2 / Y$, is sliding along the flat $z^{\alpha}$
directions [\ref{cfkn}].

We now move to the full theory, containing the $N_{TOT}$ fields $z^I
\equiv (z^{\alpha},z^{i})$. To obtain a simple expression for the
full
Ricci tensor, and inspired by the effective theories of
four-dimensional superstrings, to be discussed in the following
section, we assume some
generic scaling (homogeneity) properties for the K\"ahler potential
associated with the $z^{i}$ fields.  In a suitable parametrization,
such that one can expand for small field fluctuations around $\langle
z^{i} \rangle = 0$, we write the full K\"ahler function as
\be
\label{totkf}
\cg = - \log Y (r^{\alpha}) + \sum_A K^A_{i_A \ov{j}_A}
(r^{\alpha}) z^{i_A} \ov{z}^{\ov{j}_A}+ {1 \over 2} \sum_{A,B}
\left[ P_{i_A j_B}  (r^{\alpha}) z^{i_A} z^{j_B} + {\rm h.c.}
\right] + \log |w(z^{i})|^2  + \ldots \, ,
\ee
where $K^A_{i_A \ov{j}_A}$ is an $n_A \times n_A$ matrix and a
homogeneous function of degree $\lambda_A$, i.e.
\be
\label{homobis}
r^{\alpha} K^A_{i_A \ov{j}_A \, \alpha} = \lambda_A K^A_{i_A
\ov{j}_A} \, ,
\;\;\;\;\;
\sum_A n_A = N - n \, ,
\ee
and the dots stand for cubic or higher-order terms in the fields
$z^{i}$. To compute the coefficient $Q$ of the one-loop quadratic
divergences, we do not need to make any particular assumption about
the form of the functions~$P_{i_A j_B}$.  However, we shall see in
the following section that, in the effective theories of
four-dimensional superstrings, also the functions~$P_{i_A j_B}$
have scaling properties analogous to eq.~(\ref{homobis}),
\be
\label{homoter}
r^{\alpha} P_{i_A j_B \, \alpha} = \rho_{i_A i_B} P_{i_A j_B} \,
{}.
\ee

The K\"ahler metric associated with the full K\"ahler function of
eq.~(\ref{totkf}) has the form
\be
\label{kmetric}
\cg_{I \ov{J}} =
\left(
\begin{array}{cccc}
G_{\gamma \delta} &  &  &  \\
& \ldots &  & \\
& & K^A_{i_A \ov{j}_A} & \\
& & & \ldots
\end{array}
\right)
\, ,
\ee
and from eqs.~(\ref{homobis}) and (\ref{kmetric}) it immediately
follows that $\cg^{\alpha} \cg_{\alpha} \equiv 3$,
\be
\cg^{\alpha} (\partial_{\alpha} \partial_{\beta} \log \det G_{\gamma
\delta} ) \cg^{\beta} = 2 n \, ,
\;\;\;\;\;
\cg^{\alpha} (\partial_{\alpha} \partial_{\beta} \log \det K^A_{i_A
\ov{j}_A}
 ) \cg^{\beta} = - \lambda_A n_A \, ,
\ee
so that
\be
\cg^I R_{I \ov{J}} \cg^{\ov{J}} =
\cg^{\alpha} R_{\alpha \beta} \cg^{\beta} =
2 n - \sum_A \lambda_A n_A \, .
\ee
To include the possibility of $F_{I \ov{J}} \ne 0$, we also assume
that the gauge field metric, ${\rm Re \,}f_{ab}$, is a homogeneous
function of the variables $r^{\alpha}$ of degree $\lambda_f$, i.e.
\be
\label{lambdaf}
r^{\alpha} ({\rm Re \,} f_{ab})_{\alpha} = \lambda_f  {\rm Re
\,}f_{ab} \, .
\ee
Observe that, because of the analyticity of $f$, the only possible solutions
to eq.~(\ref{lambdaf}) correspond to $\lambda_f=0,1$, and in the latter
case $f_{ab}$ must be a linear function of the fields $z^{\alpha}$.
Denoting by $d_f$ the dimension of the gauge group, we then get
\be
\cg^I F_{I \ov{J}} \cg^{\ov{J}} =
\cg^{\alpha} F_{\alpha \beta} \cg^{\beta} =
\lambda_f d_f \, .
\ee
This allows us to rewrite the general expression for $Q$,
eq.~(\ref{qexpr}), in the final form
\be
\label{qfinal}
Q = \sum_A \left( 1+\lambda_A \right) n_A - n - \lambda_f d_f - 1 \,
{}.
\ee
{}From eq.~(\ref{qfinal}) we can
immediately read the contributions to $Q$ from the chiral and gauge
multiplets, once their scaling weights $\lambda_A$ and $\lambda_f$
are given. For example, chiral multiplets do not contribute to $Q$ if
$\lambda_A = -1$ and give a positive contribution if $\lambda_A = 0$,
whereas the $z^{\alpha}$ multiplets ($\lambda = -2$) and the massive
gauginos always give a negative contribution. We shall comment later
on the possibility of chiral multiplets with $\lambda_A < -1$, which
would provide additional negative contributions to $Q$. For the
moment,
it is important to stress again that, within our class of
supergravity
models with approximate scaling properties, requiring that
$Q=0$ amounts to
a field-independent but highly non-trivial constraint, which couples
the hidden and the observable sectors.

For the LHC models under consideration, the general supergravity mass
formulae of the previous paragraph undergo dramatic
simplifications, especially if one also assumes the scaling
properties (\ref{homoter}) for the functions $P_{i_A j_B}$.
The (non-normalized) squared mass matrix for gauginos can be written as
\be
\label{first}
\left( M_{1/2} M^{\dagger}_{1/2} \right)_{ab} = \lambda_f^2
({\rm Re \,} f)_{ab} \, m_{3/2}^2 \, .
\ee
Moving to the (non-normalized)  mass terms for the component fields
of the chiral
supermultiplets, and distinguishing between the indices $\alpha$ and
$i$, after some simple algebra we find
\be
(M_{1/2})_{\alpha \beta} = \left( - \cg_{\alpha \beta} + {1 \over 3}
\cg_{\alpha} \cg_{\beta} \right) m_{3/2} \, ,
\ee
\be
(M_0^2)_{\alpha \beta} = 0 \, ,
\ee
and
\be
\label{mugen}
\begin{array}{ccl}
\displaystyle{
(M_{1/2})_{i_A j_B}} &  =  & \displaystyle{
\left[ {w_{i_A j_B} \over w} +
P_{i_A j_B} \left( 1 + \rho_{i_A j_B} \right) \right]
m_{3/2} } \\ & & \\
& = &  \displaystyle{ w_{i_A j_B} e^{K/2} +  P_{i_A j_B} \left( 1
+ \rho_{i_A j_B} \right) m_{3/2} \, , }
\end{array}
\ee
\be
(M_0^2)_{i_A \ov{j}_B} = \delta_{AB} \left[ \left( M_{1/2}
M^{\dagger}_{1/2} \right)_{i_A \ov{j}_A} + (1 + \lambda_A)
K^A_{i_A \ov{j}_A} m_{3/2}^2 \right] \, ,
\ee
\be
\label{m3gen}
\begin{array}{c}
\displaystyle{
(M_0^2)_{i_A j_B}} =  \displaystyle{ \left[ ( 2+ \lambda_A +
\lambda_B ) {w_{i_A j_B}
\over w} +  (2+ \lambda_A + \lambda_B - \rho_{i_A j_B}) (1 +
\rho_{i_A
j_B}) P_{i_A j_B}  \right] m_{3/2}^2 }
\\ \\ =  \displaystyle{ ( 2+ \lambda_A + \lambda_B ) w_{i_A j_B}
e^{K/2} m_{3/2}
+ (2+ \lambda_A + \lambda_B - \rho_{i_A j_B}) (1 + \rho_{i_A
j_B}) P_{i_A j_B} m_{3/2}^2  \, ,
}
\end{array}
\ee
\be
\label{last}
\begin{array}{ccl}
\displaystyle{
V_{i_A j_B k_D} } & = & \displaystyle{
( 3 + \lambda_A + \lambda_B + \lambda_D )
{w_{i_A j_B k_D} \over w} m_{3/2}^2  }
\\ & & \\ & = & \displaystyle{
( 3 + \lambda_A + \lambda_B + \lambda_D )
w_{i_A j_B k_D} e^{K/2} m_{3/2} \, .}
\end{array}
\ee
A number of important consequences can be derived from
eqs.~(\ref{first})--(\ref{last}) already at this level.
Even more stringent ones will be derived in the
following section, by using additional constraints on the
functions $w$, $K^A_{i_A \ov{j}_A}$ and $P_{i_A j_B}$
coming from generalized target-space duality symmetries.

First, notice that the spin 0 fields $z^{\alpha}$ in the
supersymmetry
breaking sector have always masses ${\cal O}(m_{3/2}^2/\mpl)$, i.e.
in
the $10^{-3}$--$10^{-4} {\rm \; eV}$ range if the gravitino mass is
at the
electroweak scale, with interesting astrophysical [\ref{astro}] and
cosmological [\ref{cosmo}] implications, including a number of potential
phenomenological problems. After subtracting the goldstino,
eq.~(\ref{gold}), their spin $1/2$ partners $\chi^{\alpha}$ have all
masses equal to the gravitino mass $m_{3/2}$, as can be easily
verified by noticing that the canonically normalized mass matrix
$M_{\chi}$ is real and symmetric, and satisfies $M_{\chi}^2 = m_{3/2}
M_{\chi}$, ${\rm tr \;} M_{\chi}^2 = (n-1) m_{3/2}^2$.

Furthermore, by remembering that the chiral superfields $z^i$ should
contain the quark, lepton and Higgs superfields of the MSSM, one can
derive some predictions for the explicit mass parameters of the
MSSM. Similar predictions were derived, for special goldstino
directions and under slightly different assumptions, in
ref.~[\ref{bim}], and we agree with these results when applicable.
For the gaugino masses one finds that, if there is unification
of the gauge couplings,
$({\rm Re \,} f)_{ab} = \delta_{ab} / g_U^2$, then
\be
\label{mgaugen}
m_{1/2}^2 = \lambda_f^2 \, m_{3/2}^2 \, ,
\;\;\;\;\;
\;\;\;\;\;
(\lambda_f=0,1) \, .
\ee
As for the spin $1/2$ fermions $\chi^i$, we should distinguish two
main
possibilities. Those in chiral representations of the gauge group,
such as
the quarks and the leptons, cannot have gauge-invariant mass terms.
Those in real representations of the gauge group, such as the
Higgsino fields $\tilde{H}_1$ and $\tilde{H}_2$ of the MSSM, can have
both a `superpotential mass', proportional to $w_{i_a j_B}$, and a
`gravitational' mass, proportional to $P_{i_A j_B}$, but the
distinction between the two terms is not invariant under analytic
field redefinitions. Both these terms can in principle contribute to
the superpotential `$\mu$-term' of the MSSM, and to the associated
off-diagonal (analytic-analytic) scalar mass term $m_3^2$:
we do not give here their explicit expressions, since much simpler
ones will be obtained in the following section, within the
effective theories of four-dimensional superstrings. We anticipate
here that in these examples either the superpotential or the
gravitational contribution to $\mu$ will be present, not both.
Writing then $(M_0^2)_{i_A j_B} = (B)_{i_A j_B} (M_{1/2})_{i_A j_B}$,
in analogy with the MSSM notation, we shall find
\be
B_{H_1 H_2} = (2 + \lambda_{H_1} + \lambda_{H_2}) m_{3/2} \, ,
\ee
or
\be
\label{bstr}
B_{H_1 H_2} = (2 + \lambda_{H_1} + \lambda_{H_2} - \rho_{H_1 H_2})
m_{3/2} \, ,
\ee
respectively. Moving further to the spin 0 bosons $z^i$ in chiral
representations (squarks, sleptons, \ldots), they can only have
diagonal (analytic-antianalytic) mass terms, of the form
\be
\label{mogen}
(m_0^2)_A = (1+\lambda_A) m_{3/2}^2 \, .
\ee
{}From the previous formula, we can see, as already observed in [\ref{bim}],
that the scaling weights of the quark and lepton superfields must respect
the inequality
\be
\label{agen}
\lambda_A \ge -1 \, ,
\ee
since otherwise one would develop charge- and colour-breaking minima.
Weights smaller than $(-1)$ are allowed, instead, for the Higgs fields,
since in that case a negative contribution to $m_0^2$ can be
compensated by an extra positive contribution coming from the
$\mu$-term. Similarly, a general formula can be obtained for the
coefficients of the cubic scalar couplings of the MSSM potential,
\be
(A)_{i_A j_B k_D} = (3 + \lambda_A + \lambda_B + \lambda_D) m_{3/2}
\, .
\ee
It is remarkable that the only field dependence of the soft breaking
terms in eqs.~(\ref{mgaugen})--(\ref{agen}) is via the
gravitino mass $m_{3/2}$: this fact is welcome both to fulfil the
stringent constraints
on soft mass terms that come from flavour-changing neutral currents
[\ref{fcnc}] and to generate dynamically the hierarchy $m_{3/2} \ll
\mpl$
via MSSM quantum corrections [\ref{ekn},\ref{kpz}].

\nsect{String examples}

In this section, we apply the mass formulae obtained for the LHC
supergravity models to some concrete examples, corresponding to the
effective theories of
different four-dimensional superstring models, and to different
possible
mechanisms for spontaneous supersymmetry breaking. Our purpose is
twofold: we want not only to illustrate the previous results on
a number of representative cases, but also to justify our
assumptions, which at the pure supergravity level might appear
plausible but not really compulsory.

We have already stressed that the structure of a generic $N=1$
supergravity has a large amount of arbitrariness. Such arbitrariness
is significantly reduced if one considers the particular class of
theories that are obtained, in the low-energy limit, from some
underlying four-dimensional superstring model. Even if there are
infinitely many four-dimensional superstring vacua with unbroken
$N=1$ supersymmetry, the form of their low-energy effective theories
is
subject to important restrictions. For each of these vacua, the gauge
group and the multiplet content are uniquely specified, and  so are
the K\"ahler and the gauge kinetic functions, which, as we shall
describe in the following, do indeed exhibit the remarkable
geometrical properties assumed in the previous section. Moreover, as
an effect of the string unification of all interactions, these
theories do not contain any explicit mass parameter besides the
string
mass scale $M_S$, in the sense that all couplings and masses of the
low-energy effective theory  are associated with the VEVs of some
moduli
fields.

There is a vast literature concerning the effective supergravities
corresponding to four-dimensional superstring models with unbroken
$N=1$ local supersymmetry, both at the classical
[\ref{effcl},\ref{efferm}]
and at the quantum [\ref{effqu}] level.
The typical structure that emerges is the following. The vector
multiplets are fixed by the four-dimensional gauge group
characterizing a given class of string solutions. As for the
chiral multiplets, there is always a universal `dilaton-axion'
multiplet, $S$, singlet under the gauge group, which at the classical
level entirely determines the gauge kinetic function,
\be
\label{gkf}
f_{ab} = \delta_{ab} S \, .
\ee
Notice that, in the notation of eq.~(\ref{lambdaf}), $\lambda_f=1$ if
$S \in \{ z^{\alpha} \}$, $\lambda_f=0$ otherwise.
In addition to $S$, there are in general other singlet chiral
superfields, called `moduli', which do not appear in the
superpotential and thus correspond to classically flat directions
of the scalar potential. They parametrize the size and the shape
of the internal compactification manifold, and will be denoted here
by the generic symbols $T$ and $U$. Finally, there are other chiral
superfields, which are in general charged under the gauge group, or
at least have a potential induced by some superpotential coupling:
for the moment, we shall denote them with the generic symbol $C$,
understanding that in realistic models this class of fields should
contain the matter and Higgs fields of the MSSM.

The remarkable fact is that in the known four-dimensional string
models, in the limit where the $T$ and/or $U$ moduli are large with
respect to the string scale $M_S$, the K\"ahler manifold
for the chiral superfields obeys the properties assumed in section~2,
with well-defined scaling weights of the K\"ahler metric with respect
to
the real combinations of moduli fields $s \equiv (S+\ov{S})$, $t_i
\equiv (T_i+\ov{T}_i)$ and $u_i \equiv (U_i+\ov{U}_i)$. As we are
going to explain, these scaling weights are remnants of the
target-space duality symmetries [\ref{duality}], which survive
in the limit of large $T$ and/or $U$ moduli.
More precisely, the K\"ahler potential can be written as
\be
\label{kalstr}
K = - \log Y (s,t,u)
+ K^{(C)} (C,\ov{C};T,\ov{T};U,\ov{U}) \, .
\ee
The function $Y$ factorizes into three terms,
\be
\label{yfact}
Y = Y^{(S)} (s) \cdot Y^{(T)} (t) \cdot Y^{(U)} (u) \, ,
\ee
where
\be
Y^{(S)} = s \, ,
\ee
so that
\be
\label{scals}
s \, Y_s = Y \, .
\ee
Another general feature involves the moduli $T_i$, corresponding to
harmonic $(1,1)$ forms, associated with deformations of the K\"ahler
class of the internal compactified space. Even if their number is
model-dependent, the
fact that three of them are related to the three complex coordinates
of the internal compactification manifold implies, in the limit of
large $T$ moduli,
\be
\label{scalt}
t_i \, Y_{t_i} = 3 \, Y \, .
\ee
The moduli $U_i$ are associated with harmonic $(1,2)$ forms,
correspond to deformations of the complex structure of the internal
compactified space, and their existence, number and properties are
more model-dependent. In general, in the limit of large $U$ moduli
one can write a relation of the form
\be
\label{scalu}
u_i \, Y_{u_i} = p_U \, Y \, ,
\ee
where $p_U=0,1,2,3$ depends on the superstring model under
consideration.
Finally, keeping only quadratic fluctuations of the $C$ fields
(sufficient to evaluate the K\"ahler metric and the mass
terms around $C = \ov{C} = 0$), one can in general write
\be
\label{kmatter}
K^{(C)} = \sum_A K^A_{i_A \ov{j}_A} (t,u) C^{i_A}
\ov{C}^{\ov{j}_A} + {1 \over 2} \sum_{A,B} \left[  P_{i_A j_B}
(t,u) C^{i_A} C^{j_B} + {\rm h.c.} \right] + \ldots \, ,
\ee
with generic scaling properties of the form
\be
\label{scalct}
t_i \left( K^A_{i_A \ov{j}_A} \right)_{t_i} =
\lambda_t^A K^A_{i_A \ov{j}_A}
\, ,
\ee
\be
\label{scalcu}
u_i \left( K^A_{i_A \ov{j}_A} \right)_{u_i} =
\lambda_u^A K^A_{i_A \ov{j}_A}
\, ,
\ee
\be
\label{scalpt}
t_i (P_{i_A j_B})_{t_i} = \rho^t_{i_A j_B} P_{i_A j_B} \, ,
\ee
\be
\label{scalpu}
u_i (P_{i_A j_B})_{u_i} = \rho^u_{i_A j_B} P_{i_A j_B} \, ,
\ee
where the scaling weights of $K^A_{i_A \ov{j}_A}$ and of
$P_{i_A j_B}$ are now correlated
\be
\rho^t_{i_A j_B} = {\lambda_t^A + \lambda_t^B \over 2} \, ,
\;\;\;\;\;\;
\rho^u_{i_A j_B} = {\lambda_u^A + \lambda_u^B \over 2} \, .
\ee
As we shall see in the following examples, the fact of having definite
values for the scaling weights $\lambda_t,\lambda_u$ will amount to
significant restrictions on the possible values of the tree-level MSSM
mass parameters.

The remarkable scaling properties (\ref{scalct})--(\ref{scalpu})
follow from the discrete target-space dualities, which are
symmetries of the full K\"ahler function $\cg$. Under a generic
duality transformation, of the form
\be
z^\alpha \longrightarrow f(z^\alpha) \, ,
\ee
the K\"ahler potential transforms as
\be
K \longrightarrow K + \phi + \ov{\phi} \, ,
\ee
where $\phi$ is an analytic function of the moduli fields $z^\alpha$,
and in particular it must be that
\be
Y \longrightarrow Y e^{\phi + \ov{\phi}} \, .
\ee
Also, it is not restrictive for our purposes to consider the case in which
the fields $C_A$ transform with a specific modular weight $\lambda_A$,
\be
C_A \longrightarrow e^{- \lambda_A \phi} C_A \, .
\ee
The fact that target-space duality is a symmetry then implies a
definite transformation property for the superpotential,
\be
w \longrightarrow e^{- \phi} w \, ,
\ee
which in turn puts very strong restrictions on the superpotential
couplings, for example the cubic Yukawa couplings of the form
$h_{i_A j_B k_D} C^{i_A} C^{j_B} C^{k_D}$. If $h_{i_A j_B k_D}$
is such that, in the large moduli limit, it goes to a non-vanishing
constant (or, more generally, to a modular form of weight zero),
then we must have
\be
\lambda_A + \lambda_B + \lambda_D = 1 \, .
\ee
For example, in $Z_2 \times Z_2$ orbifolds the $h_{i_A j_B k_D}$ are
constants,
whereas in Calabi-Yau manifolds they are modular forms of weight
zero, which
approach a constant in the large volume limit for the associated
moduli.

In the case of unbroken supersymmetry, and in the large moduli limit,
the classical superpotential $w$ is independent of the $(S,T,U)$
moduli fields,
and at least quadratic in the $C$ fields. From the previous scaling
properties, it also follows that around $C=0$ one can write
\be
K^s K_s = 1 \, ,
\;\;\;\;
K^{t_i} K_{t_i} = 3 \, ,
\;\;\;\;
K^{u_i} K_{u_i} = p_U \, .
\ee
Armed with this result, we are ready to discuss spontaneous
supersymmetry breaking in the superstring effective supergravities.
As already explained, to have broken supersymmetry and vanishing vacuum
energy one needs $w \ne 0$ and $\cg^I \cg_I = 3$ at the minima of the
tree-level potential. If one takes the effective supergravities
derived from the
four-dimensional superstring models with unbroken supersymmetry, one
consistently obtains a positive semi-definite scalar potential,
admitting $C=0$ minima with unbroken supersymmetry and vanishing
vacuum energy, and flat directions along the $S$, $T$ and $U$ moduli
fields. To obtain supersymmetry-breaking minima with unbroken gauge
symmetries, i.e. the situation discussed in the previous section, one
must then introduce a superpotential modification, which generates
minima with $C=0$, $w \ne 0$, $\cg^I \cg_I = 3$ when the summation
index $I$ runs over the ($S,T,U$) moduli, $\cg^I \cg_I = 0$ when $I$
runs over the $C$ fields. This means, however, that the
superpotential modification must depend on at least some of the
($S,T,U$) moduli, since otherwise we would get,
when summing over the moduli indices, $\cg^I \cg_I = 4 + p_U$, which
would make the scalar potential strictly positive-definite and thus
not allow for the desired minima. As for the origin of possible
superpotential modifications, we must refer to the two types of
mechanisms for supersymmetry breaking considered so far in the
framework of four-dimensional string models. The first one
corresponds to exact tree-level string solutions, in which
supersymmetry is broken via orbifold compactification. The second one
is based on the assumption
that supersymmetry breaking is induced by non-perturbative phenomena,
such as gaugino condensation or something else, at the level of the
string effective field theory. These will be the two possibilities
considered in the following examples. Before moving to the examples,
we would like to present some general results for the superpotential
modification, which can be obtained as a consequence of target-space
duality.

In the case of non-perturbative supersymmetry breaking, in the absence
of a second-quantized string formalism one can assume that, at the
level of the effective supergravity, the superHiggs mechanism is induced
by a superpotential modification which preserves target-space duality
[\ref{flst}].
The relevant transformations are those acting non-trivially on the moduli
fields $z^\alpha$ associated with supersymmetry breaking. If, for example,
the modified superpotential has the form
\be
w = w_{SUSY} + A(z^\alpha) + B_{i_A j_B}(z^\alpha) C^{i_A} C^{j_B}
+ \ldots \, ,
\ee
target-space duality requires then the following transformation
properties:
\be
A(z^\alpha) \longrightarrow A(z^\alpha) e^{-\phi} \, ,
\;\;\;\;\;
B_{i_A j_B}(z^\alpha) \longrightarrow B_{i_A j_B}(z^\alpha) e^{-(1 -
\lambda_A - \lambda_B)\phi} \, .
\ee
Unfortunately, the form of the function $A(z^\alpha)$ cannot be uniquely
fixed by the requirement that it is a modular form of weight $(-1)$.
However, another important constraint comes from the physical requirement
that the potential must break supersymmetry and generate a vacuum energy at
most ${\cal O}(m_{3/2}^4)$ in the large moduli limit. This implies that
$A(z^\alpha) \longrightarrow {\rm constant} \ne 0$ for $z^\alpha
\rightarrow \infty$. This is not the case for the models of supersymmetry
breaking with minima of the effective potential at small values of $T$,
which make use of the Dedekind function $\eta(T)$ in the superpotential
modification [\ref{lust}]: either they do not break supersymmetry or they
do so with a large cosmological constant, in  contradiction with the
assumption of a constant flat background.
In the case of the function  $B_{i_A j_B}(z^\alpha)$,
it is sufficient to assume that, in the large moduli limit, $B_{i_A
j_B}(z^\alpha) \longrightarrow {\rm constant}$. For the moduli fields
that are not involved in the breaking of supersymmetry, these
asymptotic conditions are not necessary and can be relaxed.
The requirement that $A(z^\alpha) \longrightarrow {\rm constant} \ne 0$
for $z^\alpha \rightarrow \infty$ defines an approximate no-scale
model, with minima of the effective potential corresponding to
field configurations that are far away from possible $z^\alpha
\simeq {\cal O}(1)$ self-dual minima with unbroken supersymmetry
($\cg_\alpha =0$) and negative vacuum energy ${\cal O}(M_P^4)$.
Between these two classes of  extrema, there may exist other extrema
of the effective potential with $\cg_\alpha \ne 0$, but those are
generically unstable and/or have non-vanishing vacuum energy [\ref{lust}].
As for the VEVs of the moduli fields that do not contribute to supersymmetry
breaking (those with $\cg^i \cg_i = 0$), they are generically fixed to some
extended symmetry points (e.g. the self-dual points).

In the string models with tree-level supersymmetry-breaking, the
superpotential modifications in the large-moduli limit are fully
under control, since in that case the explicit form of the one-loop
string partition function is known, and one can derive the low-energy
effective theory without making any assumption. As we shall see later,
one obtains automatically the desired scaling properties of the kinetic
terms, which in some cases can produce a LHC model. In this class of
models, the large-moduli limit is a necessity, since for small values
of the moduli (close to their self-dual points) there exist Hagedorn-type
instabilities, induced by some winding modes that become tachyonic in flat
space-time [\ref{ss3},\ref{ss5}]. At the self-dual point there is a new
stable minimum with unbroken supersymmetry and negative cosmological
constant, as expected. We should stress here that the prescription for
a consistent effective field theory in the region of small moduli requires
the addition of extra degrees freedom, corresponding to the winding modes
which can become massless or tachyonic for some values of the
$T$ and/or $U$ moduli close to the self-dual points. In the large-moduli
limit, however, we can disregard the effects of these extra states and
not include them in the effective field theory. In this limit, as we shall
see, the superpotential modification associated with supersymmetry breaking
seems to violate target-space duality. On the other hand, $w_{SUSY}$ and
the K\"ahler potential mantain the same expressions as in the case of exact
supersymmetry, with the desired scaling properties that can produce a LHC
supergravity model.

\subsection{String tree-level breaking}

At the level of explicit four-dimensional $N=1$ heterotic string
constructions, the only known mechanism for spontaneous supersymmetry
breaking is the tree-level one, based on generalized coordinate-dependent
compactifications [\ref{ss1}--\ref{ss5}], which is analogous to the one
proposed by Scherk and
Schwarz for extended supergravity theories [\ref{ss}]. This mechanism
was also considered by Fayet [\ref{fayet}] and by Rohm [\ref{rohm}]
in the context of $N=2$ extended supergravity and of type-II
superstrings, respectively. In the case of $N=1$ chiral theories,
this mechanism is inconsistent at the field theory level, whereas
it can be consistently formulated in the framework of orbifold string
constructions, thanks to the existence of string `twisted states',
which contain non-trivial chiral sectors. The effective $N=1$
supergravities, corresponding to superstring models where
supersymmetry is spontaneously broken by this mechanism, were derived
in [\ref{ss2}] for fermionic constructions ($Z_2 \times Z_2$
orbifolds), and can be easily generalized to a large class of
orbifold models. The main features of these effective theories are
the following:
\begin{enumerate}
\item
The K\"ahler potential and the gauge kinetic function of the
effective theory are the same as those obtained in the limit of
unbroken supersymmetry, so that, up to analytic field redefinitions,
supersymmetry breaking is indeed
induced only by a superpotential modification.
\item
For $C=0$, the K\"ahler manifold for the $T$ and $U$ moduli can be
decomposed into the product of two factor manifolds. The first one,
described by a K\"ahler potential $K'$, involves one $T$ and
one $U$ field, to be called here $T'$ and $U'$
\be
K' = - \log [(T' + \ov{T}')(U' + \ov{U}')] \, ,
\;\;\;\;\;
(C=0) \, ,
\ee
and the second one, described by a K\"ahler potential $K''$, involves
all the remaining $T$ and $U$ moduli.
\item
The superpotential modification associated with supersymmetry
breaking does not involve the fields $S$, $T'$ and $U'$, so that
$\cg_S=K_S$, $\cg_{T'}=K_{T'}$, $\cg_{U'}=K_{U'}$. The condition
$\cg^{\alpha} \cg_{\alpha} = 3$, which must be satisfied at the
minima, is identically saturated by the fact that for $C=0$ it is
$\cg^{S} \cg_{S} = \cg^{T'} \cg_{T'} = \cg^{U'} \cg_{U'} = 1$. The
goldstino direction is then along some linear combination of the
$(S,T',U')$ fields.
\item
The superpotential modification associated with supersymmetry
breaking involves the fields appearing in $K''$, so that,
restricting the sum over $I$ to these fields, the condition
$\cg^I \cg_I = 0$ can be satisfied at all minima.
\end{enumerate}

As an illustrative example, we describe here in some detail the
models based on $Z_2 \times Z_2$ orbifolds, i.e. fermionic
constructions. In that case the K\"ahler manifold is known
[\ref{efferm}], and the K\"ahler potential reads
\be
K = K_0 + K' + K'' +  K_1 + K_2 + K_3 \, ,
\ee
where
\be
K_0 = - \log Y^{(S)} \, ,
\ee
\be
K' = - \log Y_1 \, ,
\ee
\be
K'' = - \log Y_2 - \log Y_3 \, ,
\ee
\be
K_1 = { z^{\alpha_1} \ov{z}^{\ov{\alpha}_1}  \over Y_2^{1/2}
Y_3^{1/2} } \, ,
\ee
\be
K_2 = { z^{\alpha_2} \ov{z}^{\ov{\alpha}_2}  \over Y_1^{1/2}
Y_3^{1/2} } \, ,
\ee
\be
K_3 = { z^{\alpha_3} \ov{z}^{\ov{\alpha}_3}  \over Y_1^{1/2}
Y_2^{1/2} } \, ,
\ee
and
\be
Y_1 = (T' + \ov{T}')(U' + \ov{U}') - (y^{i_1} +
\ov{y}^{\ov{i}_1})^2 \, ,
\ee
\be
Y_2 =  1 - y^{i_2} \ov{y}^{\ov{i}_2} +
\frac{1}{4} ( y^{i_2} y^{i_2} )( \ov{y}^{\ov{j}_2}
\ov{y}^{\ov{j}_2}) \, ,
\ee
\be
Y_3 =  1 - y^{i_3} \ov{y}^{\ov{i}_3} +
\frac{1}{4} ( y^{i_3} y^{i_3} )( \ov{y}^{\ov{j}_3}
\ov{y}^{\ov{j}_3}) \, .
\ee
The expressions for $K_1,K_2,K_3$ are valid only for quadratic
fluctuations around $z=0$, whereas those for $K_0,K',K"$ are
valid for arbitrary fluctuations of the associated fields.
Indeed, $Y_1$, $Y_2$ and $Y_3$ all parametrize manifolds of the
$SO(2,2+n) / [SO(2) \times SO(2+n)]$ type, with $n=n_{y_1},n_{y_2},
n_{y_3}$ respectively (here $i_{2,3} = 1, \ldots , n_{y_2,y_3} +2$),
and in principle they can
all be written in the same functional form. However, we have used
here the freedom of performing analytic field redefinitions to move
to a field basis where the superpotential assumes a particularly
simple form, reducing to a constant $k$ for $y=z=0$. Omitting
explicit indices to avoid too heavy a notation, in the chosen basis
the modified superpotential can be formally written as
\be
\label{sss}
w = k + \mu (y_2 y_2 + y_3 y_3)
+ y_1 y_2 y_3 + z_1 z_1 y_1 + z_2 z_2 y_2 + z_3 z_3 y_3
\, ,
\ee
where (in Planck units) the constants in the superpotential can be
written as
\be
k = {e_1 + e_2 \over 2} \, ,
\;\;\;\;\;
\mu={e_1 - e_2 \over 2} \, ,
\ee
and ($e_1,e_2$) are two quantized charges of order unity. One can easily
show [\ref{ss2}] that the superpotential (\ref{sss}) gives rise to a
positive-semi-definite potential, with an infinity of supersymmetry-breaking
minima at vanishing vacuum energy. Concentrating here on the minima with
$y=z=0$, the gravitino mass is
\be
\label{gravss}
m_{3/2}^2 = {k^2 \over (S+\ov{S})(T'+\ov{T}')(U'+\ov{U}')}
=  {(e_1 + e_2)^2 g_U^2 \over 2 R^2} \, ,
\ee
where $R\equiv 2 (T'+\ov{T}')^{1/2} (U'+\ov{U}')^{1/2}$ can be
interpreted as a radius in the internal space. The
right-hand side of the above formula clearly displays the so-called
decompactification problem [\ref{decomp}]: since in string models
$(e_1,e_2)$ are
quantized and of order unity, the internal
radius must be pushed to very large values, $R^{-1} \simeq 1 \tev$,
in order to have $m_{3/2} \simeq 1 \tev$.
One consequence of this fact is the existence, for all the states of
the spectrum with $R$-dependent masses, of an infinite tower of
Kaluza-Klein excitations, with masses that are integer multiples of
the gravitino mass. This fact, however, is still compatible with the
present experimental data [\ref{lrexp}]. The real problem resides in
the fact that,
in general, one expects large threshold corrections to the gauge and
Yukawa couplings, due to the contributions of the massive excitations
to the corresponding beta functions [\ref{explo}]. In the framework of
field theory, this problem has no solution.
As will be discussed later, however, in the framework of string
theory this problem can be avoided [\ref{ss4}], even if no realistic
string model with the desired features has been constructed yet.

Having the explicit form of the effective supergravity theory,
it is easy to determine the
scaling weights of the different fields with respect to $z^\alpha
\equiv (S,T',U')$,
\be
\lambda_f = 1 \, ,
\;\;
\lambda_{z_1,y_2,y_3} = 0 \, ,
\;\;
\lambda_{z_2,z_3} = -1 \, ,
\;\;
\lambda_{S,T',U',y_1} = - 2 \, ,
\ee
and to apply eq.~(\ref{qfinal}) to compute the value of $Q$,
\be
\label{qz2z2}
Q = - d_f - n_{y_1} + n_{y_2} + n_{y_3} + n_{z_1} \, .
\ee
The previous result is extremely interesting, since it shows that $Q$
can be zero if there is a relation among the number of fields
belonging to the chiral multiplets and the dimension of the gauge
group. From the pure supergravity point of view, one could always
arrange for an {\em ad hoc} cancellation by using the arbitrariness
in the choice of the gauge group and of the chiral multiplet content,
but such a solution would appear extremely unnatural. In string
models of this kind, however, this freedom is not present: in each
model, one has just to compute the resulting value of $Q$ and check
whether it is zero or not.

We now apply eqs.~(\ref{first})--(\ref{last}) to
compute the mass spectrum and comment on its most relevant features.
The gaugino masses are universal and equal to the gravitino mass: in
the standard notation of the MSSM, $m_{1/2}=m_{3/2}$. As for the
fields belonging to the chiral supermultiplets, it is important to
observe that they have mass terms coming both from the superpotential
and from the K\"ahler potential. Furthermore, one can observe that, in
the chosen parametrization and around the minima with $z=y=0$, the only
analytic bilinear terms in the $y$ and $z$ fields appearing in the
K\"ahler potential are those proportional to $y_1 y_1$, with scaling weights
\be
\rho_{y_1y_1} = -2 \, .
\ee
It is then immediate to see that, among the $z$ scalars containing
the chiral families, whose fermionic partners have all vanishing masses
around $y=z=0$, $(z_2,z_3)$ have vanishing scalar masses,
whereas $z_1$ have a universal `gravitational' mass equal to the
gravitino mass. Moving now to the $y$ fields, the $\tilde{y}_2$ and
$\tilde{y}_3$ fermions have superpotential masses equal to $\mu^2 e^K
= (e_1-e_2)^2 g_U^2 / (2 R^2)$, whereas the $\tilde{y}_1$ fermion fields
and the moduli fermions $\tilde{S}$, $\tilde{T}'$, $\tilde{U}'$ have
`gravitational' masses equal to the gravitino mass. The $y_1$ scalar
fields, which for $z=0$ belong to
the same $SO(2,n_1+2)/[SO(2)\times SO(n_1+2)]$ manifold as the $T'$
and $U'$ moduli, have vanishing diagonal (analytic-antianalytic)
masses, as a result of a cancellation between a negative and a
positive gravitational contribution. Moreover, also the off-diagonal
(analytic-analytic) masses for the $y_1$ scalars are vanishing for a
similar cancellation. The $(y_2,y_3)$ scalars have contributions to the
diagonal masses coming both from the superpotential, $\mu^2 e^K$, and
from the K\"ahler potential, $k^2 e^K$, whereas the off-diagonal
contributions come only from the superpotential and are given by
$2 \mu k e^K$. For each of the two sectors $y_2$ and $y_3$, then,
one has scalar mass eigenvalues given by $(\mu - k)^2 e^K$ and
$(\mu + k)^2 e^K$. The moduli scalars $S$,
$T'$ and $U'$ have all vanishing masses, associated to the
classically flat directions of the potential.
Finally, one can observe that the $A$ terms,
associated to the terms of the scalar potential
that are cubic in the charged
fields, are universal and given by $A=1$.

We would like to end this paragraph with some general considerations
on string models with tree-level spontaneous supersymmetry breaking,
going beyond the specific orbifold example discussed above. In these
models, $Q$ is also field-independent and given by expressions
similar to (\ref{qz2z2}). Although in field theories relations among
the dimension of the gauge group and the number of degrees of freedom
in the different scaling sectors for the chiral superfields look in
general unnatural in the absence of symmetry reasons, in string
theory such relations can be a consequence of the consistency of the
underlying superconformal symmetry and of the requirement of modular
invariance. Indeed, there exist many four-dimensional string
solutions, based on orbifold and fermionic constructions, which
exhibit spontaneously broken $N=1$ supersymmetry and a vacuum energy
scaling like $m_{3/2}^4$, with no contributions of order $m_{3/2}^2
\mpl^2$. This statement can be explicitly verified not only at the
level of the effective theory, which includes only the states that
are massless in the limit of unbroken supersymmetry, but also when
including the contributions of the massive string and
compactification modes.
This `miraculous' string cancellation can be seen as a consequence of
some hidden symmetries of the string spectrum, which imply some level
of fermion-boson mass degeneracy also in the phase with broken
supersymmetry.
As will be now discussed, this reorganization of the mass spectrum is
related to the properties of spontaneously broken $N=2$ and $N=4$
supergravities.

On orbifolds, the string partition function can be written as a sum
over different sectors, with different amounts of space-time
supersymmetry:
(i) one $N=4$ sector; (ii) one, two or three $N=2$ sectors; (iii) the
$N=1$ sectors, which in realistic models must contain the chiral
families. In all known string models with spontaneous supersymmetry
breaking at tree level,
boson-fermion mass splittings are generated in the $N=4$ sector and
in some of the $N=2$ sectors, but there are no tree-level mass
splittings in the $N=1$ sectors. This property is due to the fact
that the $N=1$ sectors are twisted ones, and thus their spectrum cannot
carry any dependence on the $T'$ and $U'$ moduli fields, associated
with the size of the internal compactification manifold. In a certain
class of models, the vanishing of the $m_{3/2}^2 \mpl^2$
contributions to the one-loop partition function follows from the
absence of $N=2$ sectors with non-zero fermion-boson mass
splittings. In that case, the only states with non-zero mass
splittings are those in the $N=4$ sector, and the absence of
$m_{3/2}^2 \mpl^2$ contributions to the vacuum energy (including also
the contributions from the string massive states) can be understood
in terms of the old result that $\str \cm^2 = 0$ in $N=4$ extended
supergravity [\ref{ss},\ref{supext}].
In other words, even if the model has only $N=1$ broken
supersymmetry, the organization of the mass splittings, both in the
light and in the heavy sectors, obeys the constraints of $N=4$
extended supersymmetry.  In principle, there is another possibility
of avoiding a non-vanishing $\str \cm^2$, which allows for the
presence of $N=2$ sectors with non-zero boson-fermion mass
splittings, at the condition of having vanishing contributions to the
$N=2$ beta functions coming from these sectors. In this case, the
vanishing of $\str \cm^2$ is a consequence of the finiteness of the
associated $N=2$ theory, and is also necessary in order to avoid
enormous threshold corrections to the string gauge coupling constant,
which would spoil the string perturbative expansion.

Before leaving this paragraph, a final comment on supersymmetry
breaking at the string tree level is in order. It is true that at
present the only known mechanism for spontaneous supersymmetry
breaking at the string tree level implies that the gravitino mass
scales like the inverse of the radius of an internal dimension, see
eq.~(\ref{gravss}). However, we cannot exclude the possibility of
constructing string models where supersymmetry is also spontaneously
broken at the tree level, but the gravitino mass has different scaling
properties with respect to the radius $R$, for example
\be
m_{3/2} = O(1) \, R^{-n} \, ,
\ee
with $n \le 3$. To obtain a gravitino mass at the 1 TeV scale, for
$n=1$ one
needs the scale associated with $R^{-1}$ to be also 1 TeV, but such a
scale can be pushed to $3 \times 10^{10} \gev$ for $n=2$ and to
$10^{13} \gev$ for $n=3$.
In the last two cases, the first Kaluza-Klein states would have
masses very much above the electroweak scale: only in the case $n=1$
they have a chance to be accessible at presently envisaged
accelerators. An argument which supports the possibility of having
models with $n=2,3$ is the following. It is known [\ref{ss2},\ref{ss5}]
that the existing models with tree-level supersymmetry breaking are
equivalent, from the point of view of the effective theory, to
particular `gaugings' of $N=4$ supergravity [\ref{gaugings}], based
on the $E_2 \in SO(6,6) \times
SU(1,1)_S$ group. However, from the field theory point of view there
are many other flat gaugings, corresponding to different subgroups
$H$ of $SO(6,6)$, for example $H=SU(2) \times SL(2,R)$,
which give rise to cases in which $n=2$ or $n=3$. Some technical
obstacles must be overcome in order to explicitly formulate string models
corresponding to these more general gaugings, and we hope to return
to this problem in the near future.

\subsection{Non-perturbative breaking}

We now discuss the other proposed mechanism for spontaneous
supersymmetry breaking in string-derived supergravity models, i.e.
the possibility of non-perturbative phenomena, which at the level of
the effective supergravity theory can again be described by a
modification of the superpotential.
In order to obtain a consistent model, with broken supersymmetry and
classically vanishing vacuum energy, the superpotential modification
must be such that $\cg^I \cg_I = 3$ around $C=0$. As already
discussed at the beginning of this section, the superpotential
modification must then contain some dependence on the $(S,T,U)$
moduli
fields in order to avoid a strictly positive potential.

For example, the simplest choice $w= k + O(C^2)$, with $w$
independent of the $(S,T,U)$ moduli, would give $\cg^I \cg_I =
K^{\hat{I}} K_{\hat{I}} = 4 + p_U$, where the index $\hat{I}$ runs
over the $(S,T,U)$ moduli, and therefore a potential around $C=0$ of
the form $V = (1+p_U) |k|^2 e^K=(1+p_U)|k|^2 / Y$.
Since the quantity at the numerator is field-independent and strictly
positive-definite, there is no stationary point for the potential,
with the exception of the boundaries of moduli space, $Y \rightarrow
\infty$, for example the decompactification limit $Y^{(T)} Y^{(U)}
\rightarrow \infty$ or the zero-coupling limit $Y^{(S)} \rightarrow
\infty$. It is then clear that, to avoid this problem, the
superpotential must depend on some of the moduli, in order to fix
some of the VEVs associated with the moduli directions.

The simplest superpotential modification follows from the conjecture
of gaugino condensation, and includes a non-trivial dependence on the
$S$ modulus. Such an assumption is made plausible by the fact that the
gauge coupling constant of the theory is determined by the VEV of the
$S$ field. An $S$-dependent superpotential modification can allow for
minima with $\cg_S = 0$, and fix the VEV of the $S$ modulus at the
minima.

Irrespectively of the details of the $S$ dependence of the
superpotential, as long as there is a field configuration of $S$
such that $\cg_S=0$, this is sufficient to create a well-behaved
positive-semi-definite potential in the absence of $U$-type moduli
($p_U=0$). When such moduli are present, one must make the further
assumption that the superpotential contains also a non-trivial
$U$-dependence, so that minima with $\cg_U=0$ can be allowed:
otherwise, the scalar potential would still remain strictly
positive-definite. Notice that the stabilization of the VEVs of the
$U$-type
moduli can be performed either at the string level, by moving to the
points of extended symmetry associated with the $U$ moduli, or at the
level of the effective theory, by extending the assumption made for
the $S$ field.

A particularly interesting scenario [\ref{moore}] for a non-perturbative
$S$-dependent superpotential is the requirement of an $SL(2,Z)$
duality, as suggested by a `dual' description of strongly coupled
strings in terms of (weakly coupled) toroidally compactified five-branes
[\ref{stdua}]. In this regime, the $T$ and $S$ duality symmetries are
interchanged, and it is consistent to treat the $T$ modulus
classically, i.e. in the field-theory limit. Then the superpotential
is
\be
w(S) = {F [J(S)] \over \eta^2(S)} \, ,
\ee
where $F$ defines a section of a flat holomorphic bundle over
$SL(2,R) / [ SO(2) \times SL(2,Z)]$, $\eta(S) \equiv e^{- \pi S / 12}
\prod_{n>0} ( 1 - e^{- 2 \pi n S})$ is the Dedekind $\eta$-function
for argument $(iS)$, and $J(S)$ is the generator of modular functions
of weight $0$. Further restrictions on $F$
occur if we impose that $w(S) \rightarrow 0$ for ${\rm Re \,} S
\rightarrow \infty$ (weakly coupled string). The potential is of the
no-scale type, $V \ge  0$, and its global minimum is at $\cg_S = 0$
[with $w(S) \ne 0$], which occurs for ${\rm Re \,} S = 1$, $T$ arbitrary.
The sliding gravitino mass is
\be
m_{3/2}^2 (T + \ov{T}) = {k \over Y^{(T)}} \, ,
\;\;\;\;\;
k = \left. {|w(S)|^2 \over Y^{(S)}} \right|_{{\rm Re} \, S=1} \,
{}.
\ee
Note that, without the $T$-flat directions, the dilaton potential
would not be positive-semi-definite, and its minimization would give
an unacceptable vacuum energy. The resolution of this puzzle is that
the sliding $T$ singlet allows for supersymmetry breaking ($\cg_T \ne
0$) with vanishing vacuum energy. On the contrary, without $T$ field,
the extremum with $\cg_S =0$ would correspond to unbroken
supergravity in anti-deSitter space.

In the following, we shall denote a superpotential
modification with non-trivial $S$ and $U$ dependence with the name of
`$T$-breaking', since in that case the condition of vanishing vacuum
energy, $\cg^I \cg_I = 3$, is saturated by the $T$ fields only. For
the models in which $p_U=3$, we may alternatively assume a
superpotential modification, which depends only on $S$ and $T$, so that
$\cg_S = \cg_{T_i} = 0$ at the minima, and the condition $\cg^I \cg_I
= 3$ is entirely saturated by the $U$ moduli: we shall call this
scenario `$U$-breaking'. In the cases in which the K\"ahler manifolds
for the $T$ and $U$ moduli are factorized, one may also consider
intermediate scenarios of $S/T/U$-breaking,
in which the superpotential modification is such that, at the minima
of the potential with $C=0$, it is identically $\cg^S \cg_S +
\cg^{T_i} \cg_{T_i} + \cg^{U_i} \cg_{U_i} = 3$, with non-vanishing
conntributions from more than one sector. We should stress again that in all
these scenarios the resulting value of $Q$, the coefficient appearing
in $\str \cm^2$, does not depend on the details of the superpotential
modification, but only on the scaling properties of the different
fields with respect to the moduli for which $\cg^{\alpha}
\cg_{\alpha} \ne 0$ (not summed) at the minima.

After these general considerations, some illustrative examples are in
order.

\begin{center}
{\bf 1. \phantom{b} Calabi-Yau manifolds}
\end{center}
We begin by recalling some generic features of the effective
supergravity theories corresponding to Calabi-Yau compactifications.
The gauge group is $E_6 \times E_8$, and the matter fields $C$ can be
divided into two groups: we shall denote by $A^{i_T \, a}$ the matter
fields in $27$ representations of $E_6$,
in one-to-one correspondence with the $T$ moduli, by $B^{i_U}_a$ the
matter fields in $\overline{27}$ representations of $E_6$, in
one-to-one correspondence with the $U$ moduli. The K\"ahler manifold
for the $T$ and $U$ moduli fields factorizes into the direct product
of two submanifolds with `special geometry' [\ref{special}].
The K\"ahler potential for
the matter fields, taking into account only quadratic fluctuations,
is
\be
\begin{array}{c}
K^{(A,B)} (A,\ov{A};B,\ov{B};T,\ov{T};U,\ov{U}) =
e^{K_2-K_1 \over 3} (K_1)_{i_T \ov{j}_T} A^{i_T \, a}
\ov{A}_a^{\ov{j}_T}
\\ \\
+
e^{K_1-K_2 \over 3} (K_2)_{i_U \ov{j}_U} B_a^{i_U} \ov{B}^{\ov{j}_U
\, a}
+
\left( P_{i_T j_U} A^{i_T \, a} B_a^{j_U}
+  {\rm h.c.} \right) \, ,
\end{array}
\ee
where $K_1 \equiv - \log Y^{(T)}$, $K_2 \equiv - \log Y^{(U)}$, and
the function $P_{i_T j_U}$ is such that [\ref{agnt}]
\be
\label{gava}
\partial_{\ov{k}_T}
\partial_{\ov{l}_U}
P_{i_T j_U} =
(K_1)_{i_T \ov{k}_T}
(K_2)_{j_U \ov{l}_U}
\, ,
\ee
which integrated gives $P_{i_T j_U} =  (K_1)_{i_T} (K_2)_{j_U} + \ldots$,
where the dots stand for terms annihilated by $\partial_{\ov{k}_T}
\partial_{\ov{l}_U}$. The situation of interest to us is the one where
the mooduli-dependence of the superpotential is such that
\be
\label{class}
\cg^s \cg_s = 3 r_s \, ,
\;\;\;\;\;
\cg^{t_i} \cg_{t_i} = 3 r_t \, ,
\;\;\;\;\;
\cg^{u_i} \cg_{u_i} = 3 r_u \, ,
\ee
with
\be
r_s + r_t + r_u = 1 \, ,
\ee
so that $\cg^I \cg_I = 3$ as desired. In this case, one can easily
rewrite eq.~(\ref{qfinal}) as
\be
\label{qpost}
Q = \left(N_{TOT}-1\right) - r_s d_f + \sum_i \left( \lambda_t^i r_t
+ \lambda_u^i r_u \right) \, ,
\ee
where now the index $i$ runs over all moduli and matter fields. The
relevant scaling weights for Calabi-Yau manifolds are the following
\be
\begin{array}{rll}
{S:} & \lambda_t = 0 \, ,  & \lambda_u=0 \, ; \\
{T:} & \lambda_t = -2 \, , & \lambda_u=0 \, ; \\
{U:} & \lambda_t = 0 \, , & \lambda_u=-2 \, ; \\
{A,B:} & \lambda_t = -1 \, , & \lambda_u=-1 \, .
\end{array}
\ee
We neglect here the possible presence of extra singlets besides the $T$
and $U$ moduli. However, their presence does not affect the quadratic
divergences if they have scaling weights $\lambda_t = \lambda_u= -1$.
The previous formulae for the scaling weights can also be used to discuss
the possibility of a gravitational contribution to the MSSM `$\mu$-term'.
In Calabi-Yau manifolds, from the general properties of the $P_{i_T j_U}$
coefficients and of the scaling weights for the matter fields (which should
contain the MSSM Higgs doublets, with the `$\mu$-term' originating from
a $27 \cdot \ov{27}$ coupling), we deduce that a non--vanishing
$\mu$ may occur when the goldstino mixes non-trivially the ($S,T,U$)
directions.

\begin{center}
{\bf 1a: Pure $T$- or pure $U$-breaking}
\end{center}
In the notation of
eq.~(\ref{yfact}), and in the limit of large $T$ moduli and small $U$
moduli, corresponding to the classical limit of Kaluza-Klein
compactifications, we can write
\be
\label{ytcy}
Y^{(T)}(t) = d_{i_T j_T k_T} t^{i_T} t^{j_T} t^{k_T} + \tau \, ,
\ee
and
\be
\label{yucy}
Y^{(U)}(u) = \sigma + \eta_{i_Uj_U} u^{i_U} u^{j_U} + \ldots \, .
\ee
For $\tau=0$, eq.~(\ref{ytcy}) corresponds to a particular solution
of
eq.~(\ref{scalt}), where $d_{ijk}$ are the (topological) intersection
matrix coefficients, and can be interpreted as a consequence of
the `special geometry' of Calabi-Yau moduli space. The constant
$\tau$ in
(\ref{ytcy}) is a perturbative correction coming from the
$\alpha'$-expansion of the associated $\sigma$-model
[\ref{candelas}].
As one can easily verify for each of the supersymmetry-breaking
mechanisms considered in the following, this correction gives rise to
harmless ${\cal O}
(m_{3/2}^4)$ contributions to the scalar potential and to equally
harmless ${\cal O} (m_{3/2}^2/\mpl^2)$ contributions to $Q$. In the
following, we shall set for simplicity $\tau=0$.

The simplest possibility is to assume a superpotential modification
that
depends on the $S$ and $U$ moduli, but not on the $T$ moduli.
Equivalently, one could directly write down the effective theory in
the large-radius limit for the K\"ahler class moduli $T$ only,
assuming that the VEVs of the complex structure moduli $U$ have
already been fixed by physics at the Planck scale, and consider a
superpotential modification that depends on the $S$ field only. One
gets a positive-semi-definite potential, thanks to the field identity
$\cg^{T_i} \cg_{T_i} \equiv 3$ and to the fact that there are values
of the $S$ and $U$ moduli satisfying the conditions $\cg_S = K_S +
(\log w)_S = 0$ and $\cg_{U_i} = K_{U_i} + (\log w)_{U_i} = 0$.
Making the identification $z^{\alpha} \equiv T_i$, one obtains
from eq.~(\ref{qpost}) that
\be
Q^{(T)} = h_{2,1} - h_{1,1} = {\chi \over 2} \, ,
\ee
where $\chi$ is the Euler characteristic of the Calabi-Yau manifold,
$h_{2,1}$ is the number of the $(2,1)$ moduli $U_i$, and $h_{1,1}$ is
the number of the $(1,1)$ moduli $T_i$. In this case, the tree-level
spectrum is particularly simple. The only massive states are the physical
$T$ fermions and the $U$ and $S$ scalars, all with masses equal to the
gravitino mass.

Because of mirror symmetry [\ref{mirror}], we can also consider the
limit specular
to the previous one, i.e. the limit of large $U$ moduli and small $T$
moduli, in which case the roles of $K_1$ and $K_2$ are interchanged,
with $d_{i_T j_T k_T} \longrightarrow d_{i_U j_U k_U}$, where $d_{i_U
j_U k_U}$ corresponds to
the classical limit of the mirror Calabi-Yau manifold. We can then
assume that the superpotential modification depends non-trivially
on the $S$ and $T$ moduli, but not on the $U$ moduli, so that
$\cg^{U_i} \cg_{U_i} \equiv 3$, due to the geometry of the $(1,2)$
moduli, and $\cg_S = \cg_{T_i} = 0$ at the minima of the potential.
Making the identification $z^{\alpha} \equiv U_i$, one obtains from
eq.~(\ref{qpost}) that
\be
Q^{(U)} = h_{1,1} - h_{2,1} = - {\chi \over 2} \, .
\ee
Specularly to the previous case, the only massive states are the
physical $U$ fermions and the $T$ and $S$ scalars, all with masses
equal to the gravitino mass.

Both for the $T$-breaking and for the $U$-breaking, the coefficient
$Q$ of the one-loop quadratic divergences is non-zero in all
physically relevant models, since in all Calabi-Yau vacua the number
of chiral fermion families is also proportional to the Euler
characteristic. We may interpret this result in the sense that,
for Calabi-Yau string solutions, pure $T$-breaking and
pure $U$-breaking are incompatible with the stability of the
hierarchy $0 \ne m_{3/2} \ll \mpl$. Notice, however, that the mirror
symmetry maps $Q$ into $-Q$: this suggests an interesting possibility,
to be described below.

\begin{center}
{\bf 1b: mixed $T/U$-breaking}
\end{center}
{}From the string point of view, as an additional possibility for
supersymmetry breaking we can consider the case in which both the $T$
and the $U$ moduli are
large, and introduce a superpotential modification such that the
potential admits minima with $\cg_{T_i} \ne 0$, $\cg_{U_i} \ne 0$,
$\cg_S=0$. The vanishing of the classical vacuum energy then implies
\be
\cg^{T_i} \cg_{T_i} = 3 \cos^2 \theta
\;\;\;\;\;\;
{\rm and}
\;\;\;\;\;\;
\cg^{U_i} \cg_{U_i} = 3 \sin^2 \theta \, .
\ee
In the notation of eq.~(\ref{class}), $r_s=0$, $r_t = \cos^2 \theta$,
$r_u = \sin^2 \theta$.
Since, as we shall see, with this mixed $T/U$-breaking it is possible
to have $Q^{(T/U)} = 0$ in Calabi-Yau models, it is appropriate to
discuss this case in some detail. First, we shall show that there are
superpotential modifications that break supersymmetry with a
positive-semi-definite potential. Neglecting the contributions of the
$C$ fields, using hatted indices for the fields
$t^{\hat{\alpha}}$ and dotted indices for the fields
$u^{\dot{\alpha}}$,
and considering for the moment a generic superpotential $w$, we can
write the scalar potential as
\be
\begin{array}{ccl}
V & = & e^K \left[ |w|^2 \cg^s \cg_s +
\left( K_{\da}w+w_{\da} \right) K^{\da\db} \left( K_{\db}w+w_{\db}
\right)
\right. \\
& & \\
& & +
\left. \left( K_{\ha}w+w_{\ha} \right) K^{\ha\hb} \left(
K_{\hb}w+w_{\hb} \right)
-
3 |w|^2
\right] \, .
\end{array}
\ee
Using the scaling properties of the K\"ahler manifolds associated
with the moduli fields $t^{\hat{\alpha}}$ and $u^{\dot{\alpha}}$,
$K^{\ha} K_{\ha} = K^{\da} K_{\da} = 3$, after some simple algebra we
can rewrite $V$ in the more suggestive form
\be
\begin{array}{ccl}
V & = & e^K \left\{ |w|^2 \cg^s \cg_s + 3
\left| w + {1 \over 3} K^{\da} w_{\da} + {1 \over 3} K^{\ha} w_{\ha}
\right|^2
+  w_{\da} \left( K^{\da\db} - {1 \over 3} K^{\da} K^{\db} \right)
\ov{w}_{\ov{\db}}
\right. \\
& & \\
& & \phantom{e^K} + \left. w_{\ha} \left( K^{\ha\hb} - {1 \over 3}
K^{\ha} K^{\hb} \right) \ov{w}_{\ov{\hb}} - {1 \over 3} \left[ \left(
K^{\da} w_{\da} \right) \left( K^{\ha} \ov{w}_{\ov{\ha}} \right) + {\rm
h.c.} \right] \right\} \, .
\end{array}
\ee
For a generic superpotential $w$, the above scalar potential is not
manifestly positive-semi-definite. However, we shall now show that
there exists a generic class of superpotential modifications that
ensure this desired property. Consider first an $S$-dependence of $w$
such that the equation $\cg_S=0$ can have a solution at some finite
value of $S$. Then assume that $w$ depends only on a linear
combination of the $T^{\ha}$ and the $U^{\da}$ moduli,
\be
w = w \left( d_{\ha} T^{\ha} + d_{\da} U^{\da} \right) \, .
\ee
With the above choice of superpotential, the scalar potential becomes
\be
\label{vcy}
\begin{array}{ccl}
V & = & e^K \left\{ 3
\left| w + {1 \over 3} K^{\da} d_{\da} w' + {1 \over 3} K^{\ha}
d_{\ha} w' \right|^2 +  |w'|^2 d_{\da}  \left( K^{\da\db}
- {1 \over 3} K^{\da} K^{\db} \right) \ov{d}_{\db}
\right. \\
& & \\
& & + \left. |w'|^2 d_{\ha}  \left( K^{\ha\hb} - {1 \over
3} K^{\ha} K^{\hb} \right) \ov{d}_{\hb} - {1 \over 3} |w'|^2 \left[
\left( K^{\da} d_{\da} \right) \left( K^{\ha} \ov{d}_{\ha} \right) +
{\rm h.c.} \right] \right\} \, .
\end{array}
\ee
Since, as already noticed, $K^{\ha} = - t^{\ha}$ and $K^{\da} = - u^{\da}$,
the last term of the previous
equation, which is not manifestly positive-semi-definite, identically
vanishes if the $d_{\ha}$ are purely real and the $d_{\da}$
purely imaginary (or, more generally, whenever their relative phase
is equal to $e^{i \pi/2}$). The minima with broken supersymmetry are
those for which both $w$ and $w'$ are different from zero. On the
other hand, having vanishing vacuum energy at the minima implies
\be
d_{\da}  \left( K^{\da\db} - {1 \over 3} K^{\da} K^{\db} \right)
\ov{d}_{\db}
=
d_{\ha}  \left( K^{\ha\hb} - {1 \over 3} K^{\ha} K^{\hb} \right)
\ov{d}_{\hb}
=
0 \, .
\ee
It is clear that the required conditions can be simultaneously
satisfied, since both the effective metrics
$K^{\da\db}-(1/3)K^{\da}K^{\db}$ and $K^{\ha\hb} -
(1/3)K^{\ha}K^{\hb}$ are positive semi-definite and
have always one zero eigenvalue, as a consequence
of the field identitities $K^{\ha} K_{\ha} = K^{\da} K_{\da} = 3$. In
the case in which there are only one $T$ and one $U$ moduli, the second
and the third terms in the potential $(\ref{vcy})$ are identically
vanishing due to the above identitities, and  one is just left with
the first positive-semi-definite term. In the case of many $T$ and $U$
moduli, the minima of the potential must correspond to configurations
of the $T$ and $U$ moduli such that
\be
\left( K^{\ha\hb} - {1 \over 3} K^{\ha} K^{\hb} \right) \ov{d}_{\hb}
=
\left( K^{\da\db} - {1 \over 3} K^{\da} K^{\db} \right) \ov{d}_{\db}
=
0 \, .
\ee
Since $K^{\ha}$ and $K^{\da}$ are linear in $t^{\ha}$ and $u^{\da}$,
respectively, it is convenient to perform the field redefinitions
\be
T' \equiv d_{\ha} T^{\ha}
\;\;\;\;\;
{\rm and}
\;\;\;\;\;
U' \equiv - i d_{\da} U^{\da} \, .
\ee
In terms of the redefined fields $T'$ and $U'$, the potential
(\ref{vcy})
becomes
\be
V = 3 \left| w - {1 \over 3} t' w_{t'} - {i \over 3} u' w_{u'}
\right|^2 \, ,
\ee
which is manifestly positive-semi-definite, and vanishes whenever
\be
\label{mincy}
w = {1 \over 3} t' w_{t'} + {i \over 3} u' w_{u'} \, ,
\ee
with $w = w(T' + i U')$. Using the minimization condition
(\ref{mincy}), and disregarding at first the matter fields,
one finds
\be
\cg^{T_i} \cg_{T_i} \equiv \cg^{\ha} \cg_{\ha} = 3 \cos^2 \theta \, ,
\;\;\;\;\;
\cg^{U_i} \cg_{U_i} \equiv \cg^{\da} \cg_{\da} = 3 \sin^2 \theta \, ,
\ee
where
\be
\cos^2 \theta \equiv {u'^2 \over t'^2 + u'^2}
\, ,
\;\;\;\;\;
\sin^2 \theta \equiv {t'^2 \over t'^2 + u'^2}
\, ,
\ee
so that $\cg^{T_i} \cg_{T_i}+\cg^{U_i} \cg_{U_i} \equiv  \cg^{\ha}
\cg_{\ha} + \cg^{\da} \cg_{\da} \equiv 3$ for any value of $\theta$.
The directions $t' \equiv (T'+\ov{T}')$ and $u' \equiv (U'+\ov{U}')$
are arbitrary, while $(T-\ov{T})$ and $(U-\ov{U})$ are fixed in terms
of $t'$ and $u'$. Once the direction of the goldstino has been specified,
the computation of the coefficient of quadratic divergences is a
straightforward application of eq.~(\ref{qpost}), and one gets
\be
\label{qtu}
Q^{(T/U)} = (h_{2,1}-h_{1,1}) \cos 2 \theta =  {\cos
2 \theta \over 2} \chi  \, .
\ee
As a result, this type of breaking allows for $Q^{(T/U)}=0$ in the
direction associated with $\theta = \pi/4$, corresponding to the case
in which the $T$ and $U$ moduli contribute to supersymmetry breaking
with equal strength, $\cg^{T_i} \cg_{T_i} = \cg^{U_i} \cg_{U_i} = 3/2$.
Despite the presence of $T/U$ mixing, the fact that the charged scalars
have all weight (-1) implies that both their soft masses and the
associated $\mu$ term are classically vanishing.

Even though we concentrated here on a specific example, eq.~(\ref{qtu})
can be easily extended to all string-derived supergravity models in
which, at the supersymmetry breaking minima of the
potential, $\cg^S \cg_S=0$, $\cg^{T_i} \cg_{T_i} + \cg^{U_i} \cg_{U_i}
= 3$, and there is at least one field direction, of the form $a_{\ha}
t^{\ha} + b_{\da} u^{\da}$, with $a$ and $b$ arbitrary vectors, in
which the classical potential is flat.

As a final comment, we observe that the considerations made here for
Calabi-Yau manifolds can be easily extended to arbitrary symmetric
orbifolds, whose effective theories are known [\ref{effcl}].
Without going into details, consider for example the case
of full $T$-breaking, so that $z^{\alpha} = T_i$. In this case
the scaling weights are known, and can be summarized as follows:
$\lambda_T = - 2$ for the $T_i$ moduli; $\lambda_F = - 1$
for the untwisted families, in $27$ representations of $E_6$;
for the twisted moduli, $\lambda_{TM}=-3$ when there are no $N=2$
sectors, $\lambda_{TM} = - 2$ in the presence of $N=2$ sectors;
for the twisted families, $\lambda_{TF}=-2$ when there are no $N=2$
sectors, $\lambda_{TF} = - 1$ in the presence of $N=2$ sectors.
The fact that one can have twisted families with $\lambda_{TF}=-2$
implies that the generic symmetric orbifolds, with the only exception
of the $Z_2 \times Z_2$ case, are not compatible with full $T$-breaking.

\begin{center}
{\bf 2. \phantom{b} $Z_2 \times Z_2$ orbifolds}
\end{center}
In these models, the K\"ahler potential is identical to the one
discussed in the string example of the previous paragraph, but we
prefer here to perform an analytic field redefinition, in order to
write $Y_1$, $Y_2$ and $Y_3$ all in the same functional form
\be
Y_A = (T_A + \ov{T}_A)(U_A + \ov{U}_A) - (y^{i_A} +
\ov{y}^{\ov{i}_A})^2 \, ,
\;\;\;\;
(A=1,2,3) \, .
\ee

For $z=0$, and for an arbitrary superpotential $w$, the scalar potential
of the models under consideration reads [\ref{fkpz1}]
\be
V = V_0 + \sum_{A=1}^3 V_A + V_D \, ,
\ee
where
\be
V_0 = e^K \left| w - (S + \ov{S}) w_S \right|^2 \, ,
\ee
and
\bea
V_A  & = &  e^K \left[ \left| w - w_{T_A} (T_A + \ov{T}_A) -
w_{U_A} (U_A + \ov{U}_A)  \right. \right. \nonumber \\
& & \left. \left. - w_{i_A} (y_{i_A}  + \ov{y}_{i_A}) \right|^2 +
Y_A \left( {| w_{i_A} |^2 \over 2} - \ov{w}_{\ov{T}_A} w_{U_A} -
\ov{w}_{\ov{U}_A} w_{T_A} \right) \right]
\, .
\eea
One can easily extend the calculation to include the dependence on
the $z$ fields [\ref{fkpz1}], but this is not necessary for the following
considerations.
Notice that the last term in $V_1,V_2,V_3$ is not manifestly positive
semi-definite. In the following, however, we shall consider superpotential
modifications such that the last term vanishes and supersymmetry breaking
minima with vanishing vacuum energy are generated.

\begin{center}
{\bf 2a: $(T_1,T_2,T_3)$-breaking}
\end{center}
Consider a superpotential modification with a non-trivial dependence on
$S$ and on the moduli $(U_1,U_2,U_3)$, but no dependence on the moduli
($T_1,T_2,T_3$). Then, if $\cg_S = \cg_{U_A} = 0 $ is allowed in
configuration space, supersymmetry is broken with $V=0$ and $z^{\alpha}
\equiv (T_1,T_2,T_3)$. The scaling weights of the different fields are
\be
\lambda_f = \lambda_{S,U_A} = 0 \, ,
\;\;\;
\lambda_{y_A,z_A} = -1 \, ,
\;\;\;
\lambda_{T_A} = - 2 \, .
\ee
The interesting result is that in all $Z_2 \times Z_2$ models of this
type one finds~[\ref{fkpz1}]
\be
Q = 0 \, .
\ee
As for the particle spectrum, after observing that $\rho_{y_1y_1}=
\rho_{y_2y_2}=\rho_{y_3y_3}=-1$, one finds that, besides possible
superpotential masses for the $S$ and $U$ superfields, the only states with
non-vanishing supersymmetry breaking masses, all identical to the
gravitino mass $m_{3/2}$, are the physical $T$ fermions and the $S$
and $U$ bosons.

A completely symmetric result holds if one assumes full $U$-breaking instead of
full $T$-breaking.

\begin{center}
{\bf 2b: $(T_1,U_1,T_2)$-breaking}
\end{center}
We assume here, for the sake of discussion, the existence of a
superpotential modification that generates a positive-semi-definite
potential, with supersymmetry-breaking minima such that $z^{\alpha}
\equiv (T_1,U_1,T_2)$. In such case, the gravitino mass would scale
like $R^{-2}$. The scaling weights of the different fields would be
\be
\lambda_f = \lambda_{S,T_3,U_2,U_3,y_3} = 0 \, ,
\;\;
\lambda_{z_3} = - {3 \over 2} \, ,
\;\;
\lambda_{y_2,z_2} = -1 \, ,
\;\;
\lambda_{z_1} = - {1 \over 2} \, ,
\;\;
\lambda_{T_1,U_1,T_2,y_1} = - 2 \, ,
\ee
and the coefficient of the quadratic divergences would read
\be
Q = - n_{y_1} + n_{y_3} + {1 \over 2} n_{z_1} - {1 \over 2}
n_{z_3} \, .
\ee
Notice, however, that this type of breaking can only be consistent
if $n_{z_3} = 0$, since for the fields $z_A$, which transform in
spinorial representations of the unbroken orthogonal gauge group,
and therefore cannot have gravitational mass terms,
it is not permitted to have $\lambda_{z_A} < - 1$, which would
generate a negative squared mass for the associated scalars.  One can
verify that models with the latter property can be formulated via fermionic
string constructions. On the other hand, the vanishing of $Q$ would not
be automatic, but would require a non-trivial relation among the field
representations.

\begin{center}
{\bf 2c: $(S/T_2,T_1,T_3)$-breaking}
\end{center}
Another interesting situation occurs when the $S$ moduli and one of the
compactification moduli are mixed, in a way similar to the way the $T/U$
mixing was occurring in the previous Calabi-Yau example.

Imagine a superpotential of the form
\be
w = w_{SUSY} + w_0 (T_2+iS,U_A) \, ,
\ee
with no explicit dependence on the moduli $T_2-iS$, $T_1$, $T_3$.
One can explicitly verify that the scalar potential is positive
semi-definite and, provided that the configurations with $\cg_U
=\cg_y = \cg_z = 0$ are allowed, it admits supersymmetry breaking
minima with
\be
\cg^S \cg_S = \sin^2 \theta \, ,
\;\;\;\;\;
\cg^{T_2} \cg_{T_2} = \cos^2 \theta \, ,
\;\;\;\;\;
\cg^{T_1} \cg_{T_1} = \cg^{T_3} \cg_{T_3} = 1 \, ,
\ee
where
\be
\sin^2 \theta = {t_2^2 \over t_2^2 + s^2}
\ee
corresponds to a flat direction. In this case, one finds
\be
\label{st1}
Q = \sin^2 \theta \left( {n_{z_1} + n_{z_3} \over 2} + n_{y_2} - d_f \right)
\, ,
\ee
and for $\sin^2 \theta = 0$ one recovers the result of pure $T$-breaking.
However, one can have $Q=0$ also when $\sin^2 \theta \ne 0$ but its
coefficient in eq.~(\ref{st1}) vanishes, which implies a non-trivial
relation among the field representations. The latter case would allow for
non-vanishing gaugino masses $m_{1/2}^2 = \sin^2 \theta \, m_{3/2}^2$.

\begin{center}
{\bf 2d: $(S/T_2,T_1,U_1)$-breaking}
\end{center}
Imagine finally a superpotential modification that generates a
positive-semi-definite potential, with supersymmetry-breaking minima
corresponding to
\be
\cg^S \cg_S = \sin^2 \theta \, ,
\;\;\;\;\;
\cg^{T_2} \cg_{T_2} = \cos^2 \theta \, ,
\;\;\;\;\;
\cg^{T_1} \cg_{T_1} = \cg^{U_1} \cg_{U_1} = 1 \, ,
\ee
where
\be
\sin^2 \theta = {t_2^2 \over t_2^2 + s^2}
\ee
corresponds to a flat direction. In this case, one would find
\be
Q = \sin^2 \theta \left( - d_f + n_{y_2} + {n_{z_1} + n_{z_3} \over 2}
\right) + \left( n_{y_3} - n_{y_1} + {n_{z_1} - n_{z_3} \over 2} \right)
\, .
\ee
For $\sin^2 \theta = 1$ we recover the result of the string tree-level
breaking, for $\sin^2 \theta = 0$ the result of $(T_1,U_1,T_2)$ breaking.

\nsect{Conclusions and outlook}

In this paper we have argued that, in realistic models of
spontaneously broken supergravity, the desired hierarchy $m_Z,m_{3/2}
\ll \mpl$ can be stable, and eventually find a natural dynamical
explanation, when quantum loop corrections to the effective
potential do not contain terms quadratic in the cut-off scale,
controlled at the one-loop level by $Q$, as defined in
eqs.~(\ref{genstr})--(\ref{ficci}). Requiring broken supersymmetry
with vanishing vacuum energy and vanishing $Q$ (modulo corrections
suppressed by $m_{3/2}^2/\mpl^2$ or exponentially) defines a highly
non-trivial constraint on the K\"ahler potential $K$ and the gauge
kinetic function $f_{ab}$, including both the observable and the
hidden sectors of the theory, as well as on the mechanism for
spontaneous supersymmetry breaking. In the presence of some
approximate scaling properties of the gauge and K\"ahler metrics,
with respect to the fields with non-vanishing components in the
direction of the goldstino, the contributions to $Q$ of the different
degrees of freedom that get mass via supersymmetry breaking depend
only on their scaling weights $\lambda_i$, and not on the VEVs of the
sliding singlet fields in the
hidden sector. We have derived a similar result for the individual
mass matrices of the theory, and in particular for the explicit mass
parameters of the MSSM, $(m_{1/2}, m_0,A, \mu, m_3^2)$, which take
very simple expressions in terms of the assumed scaling weights.
These expressions for $Q$ and for the mass parameters of the MSSM
find
a deeper justification in the effective theories of four-dimensional
superstrings, where supersymmetry breaking is described either at the
string tree-level or, by assuming some non-perturbative phenomena,
only in the effective field theory. In these theories, the full
particle content and
the approximate scaling weights are completely fixed. The origin of
the
approximate scaling properties of the superstring effective theories
is
due to target-space modular invariance, and the scaling weights
$\lambda_i$ are nothing but the target-space duality weights with
respect to the moduli fields, which participate in the
supersymmetry-breaking mechanism.
Indeed, in the limit of large moduli the discrete
target-space duality symmetries are promoted to some accidental
scaling symmetries of the gauge and matter kinetic terms in the
effective supergravity theory. As an application, we gave explicit
expressions for $Q$ and for the MSSM mass terms in the effective
theories of some representative four-dimensional superstring models.

When one changes the direction of the goldstino in the space of the moduli
fields, the value of $Q$ changes accordingly, in a very simple way. As an
example, we discussed the case of mixed $T/U$-breaking in Calabi-Yau
compactifications, and we showed that the case of diagonal breaking,
$\cg^{T_i} \cg_{T_i} = \cg^{U_i} \cg_{U_i} = 3/2$, corresponds to
vanishing $Q$.
In the effective theories of fermionic constructions ($Z_2 \times Z_2$
asymmetric orbifolds), the case of full $T$-breaking, $\cg^{T_A} \cg_{T_A}
=3$, also corresponds to vanishing $Q$, but one can also conceive cases
of mixed $(S,T,U)$ breaking in which $Q=0$ may be realized. As for the
explicit string
constructions in which supersymmetry is spontaneously broken at the
tree level, all the presently known solutions have $\cg^T \cg_T =
\cg^U \cg_U  = \cg^S \cg_S = 1$, and in some of them the constraint
$Q=0$ is satisfied, due to some accidental organization of the
massive string spectrum, which is very similar to the one of spontaneously
broken $N=4$ extended supergravities.
Thus we have identified in $Q=0$ another criterion for a
consistent choice of the supersymmetry breaking directions, $\{
z^{\alpha} \}$ such that $\cg^\alpha \cg_\alpha \ne 0$, in the
string-induced effective supergravities.

Some comments on our results are in order, so that we can better
identify their interpretation. The first one concerns the contribution
to $Q$ of the massive string modes. In the case of string tree-level
breaking, one can easily compute such a contribution
directly at the string level, since one knows
the explicit form of the one-loop string partition function. One can
identify a class of physically relevant four-dimensional orbifold
models where the result is the desired one,
\be
V_{eff} = {\rm (constant)} \times m_{3/2}^4 + \ldots \, ,
\ee
where the constant depends on the number of residual bosonic and
fermionic massless string states after supersymmetry breaking, and
the dots represent the contributions of the massive string states,
which, in the large moduli limit (large
compactification radius $R$ with respect to the string scale
$\alpha'^{1/2}$), are exponentially suppressed, as
$e^{-c R^2/\alpha'}$. The validity of the above behaviour follows
from two facts: (i) the absence of the $N=2$ sector associated with
the $R$-dependent threshold corrections; (ii) the fact that the
contributions to the vacuum energy come only from the $N=4$ sector,
where, including also the string massive states, one still has the
well-known sum rules of $N=4$ supergravity,
$\str \cm^n = 0$ for $n < 4$ and $\str \cm^4 = {\rm (constant)}
\times m_{3/2}^4$. This result explicitly displays two remarkable
properties. The first is the
exponential suppression of the infinitely many massive string states
in the large moduli limit: the contribution to $Q$, and also to the
terms of the one-loop vacuum energy
${\cal O}(m_{3/2}^4)$, is entirely given by the one calculated in the
effective supergravity theory, in analogy to what happens when
computing
string threshold corrections to gauge and Yukawa couplings. This fact
suggests the possibility that, at least for string
tree-level breaking, the coefficient $Q$ is a topological number,
calculable from the effective field theory data only, without the
need of knowing all the massive string modes.
If this conjecture turns out to be true, then it is not inconceivable
that $Q$, in analogy with anomalies, is a purely one-loop phenomenon:
if so, the hierarchy problem could be solved in the class of models where
$Q=0$. This conjecture would be much more difficult to prove in the case
of non-perturbative breaking at the field-theory level.
Still, there are some indications that $Q$ could be related to an
anomaly coefficient constructed from some composite currents.
We hope to return to this point in a future publication.

Among the models we have considered, the more realistic and
constrained ones appear to be those where $\cg_S \ne 0$,
since they exhibit a simple tree-level generation of
the scalar and gaugino mass terms, as well as of the $\mu$-term,
for the MSSM. However, the cancellation of quadratic divergences
requires in these models an interplay between real and chiral representations
of the gauge group. Moreover, with $\cg_S \ne 0$ the flatness of the
potential along some directions requires a specific mixing with some
other moduli, to obtain vanishing tree-level cosmological constant with
sliding gravitino mass. In this spirit, dilaton dominance [\ref{blm}]
seems difficult to reconcile with a sliding gravitino mass.

Since in this paper we looked at tree-level mass formulae and at
one-loop quadratic divergences, we did not consider string loop
corrections to the effective supergravity theory, which usually modify
the K\"ahler potential and the gauge couplings. In the framework of a
consistent perturbative expansion in the string coupling constant, these
corrections are in general negligible in first approximation. As observed
in [\ref{bim}], however, they could be relevant in the cases in which some
physical parameters are accidentally vanishing at the tree level.

One of the assumptions of the present work was the alignment of the
goldstino along directions that are gauge singlets at the Planck scale.
This implies, in particular, the existence of scalar particles with
interactions of gravitational strength and masses of order $m_{3/2}^2
/\mpl$, with interesting astrophysical and cosmological implications,
including a number of potential phenomenological problems. More general
situations may however be possible, when some underlying symmetry controls
the hidden sector physics [\ref{rhs}]. It is still an open question
whether in these models one can have a naturally vanishing cosmological
constant, modulo ${\cal O}(m_{3/2}^4)$ corrections, and at the same time
guarantee the stability of the scale hierarchies.

\section*{Acknowledgements}
One of us (F.Z.) would like to thank the ITP at Santa Barbara and the
Physics Department at UCLA for the kind hospitality during the final
stage of this work, and J.~Bagger and L.~Randall for useful discussions.

\newpage

\section*{References}
\begin{enumerate}
\item
\label{hierarchy}
K.~Wilson, as quoted by L.~Susskind, Phys. Rev. D20 (1979) 2019;
\\
E.~Gildener and S.~Weinberg, Phys. Rev. D13 (1976) 3333;
\\
S.~Weinberg, Phys. Lett. 82B (1979) 387;
\\
G.~'t~Hooft, in {\em Recent developments in gauge theories},
G.~'t~Hooft et al., eds., Plenum Press, New York, 1980, p.~135.
\item
\label{les}
L. Maiani, Proc. Summer School on Particle Physics, Gif-sur-Yvette,
1979
(IN2P3, Paris, 1980), p. 1;
\\
M. Veltman, Acta Phys. Polon. B12 (1981) 437;
\\
E. Witten, Nucl. Phys. B188 (1981) 513;
\\
See also:
\\
S. Weinberg, as in ref.~[\ref{hierarchy}].
\item
\label{mssm}
P.~Fayet, Phys. Lett. B69 (1977) 489 and B84 (1979) 416;
\\
S.~Dimopoulos and H.~Georgi, Nucl. Phys. B193 (1981) 150.
\\
For reviews and further references, see, e.g.:
\\
H.-P. Nilles, Phys. Rep. 110 (1984) 1;
\\
S. Ferrara, ed., `Supersymmetry' (North-Holland, Amsterdam, 1987);
\\
F.~Zwirner, in `Proceedings of the 1991 Summer School in High Energy
Physics and Cosmology', Trieste, 17 June--9 August 1991 (E.~Gava,
K.~Narain, S.~Randjbar-Daemi, E.~Sezgin and Q.~Shafi, eds.), Vol.~1,
p.~193.
\item
\label{cww}
S.~Coleman and E.~Weinberg, Phys. Rev. D7 (1973) 1888;
\\
S.~Weinberg, Phys. Rev. D7 (1973) 2887;
\\
J.~Iliopoulos, C.~Itzykson and A.~Martin, Rev. Mod. Phys. 47 (1975) 165.
\item
\label{veltman}
M.~Veltman, as in ref.~[\ref{les}].
\item
\label{einhorn}
M.B.~Einhorn and D.R.T.~Jones, Phys. Rev. D46 (1992) 5206.
\item
\label{zumino}
B.~Zumino, Nucl. Phys. B89 (1975) 535.
\item
\label{fgp}
S.~Ferrara, L.~Girardello and F.~Palumbo, Phys. Rev. D20 (1979) 403.
\item
\label{gg}
L.~Girardello and M.T.~Grisaru, Nucl. Phys. B194 (1982) 65.
\item
\label{cfgvp}
E.~Cremmer, S.~Ferrara, L.~Girardello and A.~Van~Proeyen,
Nucl. Phys. B212 (1983) 413;
\\
J.~Bagger, Nucl. Phys. B211 (1983) 302.
\item
\label{bfs}
A.H.~Chamseddine, R.~Arnowitt and P.~Nath, Phys. Rev. Lett. 49 (1982) 970;
\\
R.~Barbieri, S.~Ferrara and C.A.~Savoy, Phys. Lett. B119 (1982) 343;
\\
E.~Cremmer, P.~Fayet and L.~Girardello, Phys. Lett. B122 (1983) 41;
\\
L.~Hall, J.~Lykken and S.~Weinberg, Phys. Rev. D27 (1983) 2359;
\\
S.K.~Soni and H.A.~Weldon, Phys. Lett. B126 (1983) 215;
\\
G.F.~Giudice and A.~Masiero, Phys. Lett. B206 (1988) 480.
\item
\label{weinberg}
S.~Coleman and F.~de Luccia, Phys. Rev. D21 (1980) 3305;
\\
S.~Weinberg, Phys. Rev. Lett. 48 (1982) 1776.
\item
\label{bc}
R.~Barbieri and S.~Cecotti, Z.~Phys C17 (1983) 183;
\\
M.K.~Gaillard and V.~Jain, Phys. Rev. D49 (1994) 1951.
\item
\label{grk}
M.T.~Grisaru, M.~Rocek and A.~Karlhede, Phys. Lett. B120 (1983) 110.
\item
\label{destab}
J.~Polchinski and L.~Susskind, Phys. Rev. D26 (1982) 3661;
\\
H.-P.~Nilles, M.~Srednicki and D.~Wyler, Phys. Lett. B124 (1983) 337;
\\
A.B.~Lahanas, Phys. Lett. B124 (1983) 341;
\\
U.~Ellwanger, Phys. Lett. B133 (1983) 187;
\\
J.~Ellis, C.~Kounnas and D.V.~Nanopoulos, Nucl. Phys. B241 (1984) 406;
\\
J.~Polchinski, preprint UTTG-14-85 (1985), published in the Proceedings
of the 6th Workshop on Grand Unification, Minneapolis,  April 18--20,
1985;
\\
J.~Bagger and E.~Poppitz, Phys. Rev. Lett. 71 (1993) 2380.
\item
\label{fds}
P.~Candelas, G.~Horowitz, A.~Strominger and E.~Witten, Nucl. Phys.
B258 (1985) 46;
\\
L.~Dixon, J.~Harvey, C.~Vafa and E.~Witten, Nucl. Phys. B261 (1985)
678;
\\
K.S.~Narain, Phys. Lett. B169 (1986) 41;
\\
K.S.~Narain, M.H.~Sarmadi and E.~Witten, Nucl. Phys. B279 (1987) 369;
\\
W.~Lerche, D.~L\"ust and A.N.~Schellekens, Nucl. Phys. B287 (1987)
477;
\\
H.~Kawai, D.C.~Lewellen and S.-H.H. Tye, Nucl. Phys. B288 (1987) 1;
\\
K.S.~Narain, M.H.~Sarmadi and C.~Vafa, Nucl. Phys. B288 (1987) 551;
\\
I.~Antoniadis, C.~Bachas and C.~Kounnas, Nucl. Phys. B289 (1987) 87.
\\
For a review and further references see, e.g.:
\\
B.~Schellekens, ed., {\em Superstring construction} (North-Holland,
Amsterdam, 1989).
\item
\label{fkpz1}
S.~Ferrara, C.~Kounnas, M.~Porrati and F.~Zwirner,
Phys. Lett. B194 (1987) 366.
\item
\label{cfkn}
E.~Cremmer, S.~Ferrara, C.~Kounnas and D.V.~Nanopoulos,
Phys. Lett. B133 (1983) 61.
\item
\label{ekn}
J.~Ellis, A.B.~Lahanas, D.V.~Nanopoulos and K.~Tamvakis, Phys. Lett.
B134 (1984) 429;
\\
J.~Ellis, C.~Kounnas and D.V.~Nanopoulos,
Nucl. Phys. B241 (1984) 406 and B247 (1984) 373.
\item
\label{savoy}
N.~Dragon, U.~Ellwanger and M.G.~Schmidt, Nucl. Phys. B255 (1985) 549
and Prog. in Part. and Nucl. Phys. 18 (1987) 1;
\\
S.P.~Li, R.~Peschanski and C.A.~Savoy, Phys. Lett. B178 (1986) 193,
Nucl. Phys. B289 (1987) 206 and Phys. Lett. B194 (1987) 226.
\item
\label{softstr}
A.~Font, L.E.~Ib\`a\~nez, D.~L\"ust and F.~Quevedo, Phys. Lett.
B245 (1990) 401;
\\
M.~Cvetic, A.~Font, L.E.~Ib\`a\~nez, D.~L\"ust and F.~Quevedo,
Nucl. Phys. B361 (1991) 194;
\\
L.E.~Ib\`a\~nez and D.~L\"ust, Nucl. Phys. B382 (1992) 305;
\\
B.~de Carlos, J.A.~Casas and C.~Mu\~noz, Phys.Lett. B299 (1993) 234;
\\
V.S.~Kaplunovsky and J.~Louis, Phys. Lett. B306 (1993) 269;
\\
R.~Barbieri, J.~Louis and M.~Moretti, Phys. Lett. B312 (1993) 451.
\item
\label{bim}
A.~Brignole, L.E.~Ib\`a\~nez and C.~Mu\~noz, Madrid preprint
FTUAM-26/93.
\item
\label{effcl}
E.~Witten, Phys. Lett. B155 (1985) 151;
\\
A.~Strominger, Phys. Rev. Lett. 55 (1985) 2547;
\\
A.~Strominger and E.~Witten, Comm. Math. Phys. 101 (1985) 341;
\\
S.~Ferrara, C.~Kounnas and M.~Porrati, Phys. Lett. B181 (1986) 263;
\\
C.~Burgess, A.~Font and F.~Quevedo, Nucl. Phys. B272 (1986) 661;
\\
N.~Seiberg, Nucl. Phys. B303 (1988) 286;
\\
M.~Cvetic, J.~Louis and B.~Ovrut, Phys. Lett. B206 (1988) 227;
\\
S.~Ferrara and M.~Porrati, Phys. Lett. B216 (1989) 1140;
\\
M.~Cvetic, J.~Molera and B.~Ovrut, Phys. rev. D40 (1989) 1140;
\\
L.~Dixon, V.~Kaplunovsky and J.~Louis, Nucl. Phys. B329 (1990) 27.
\item
\label{efferm}
S.~Ferrara, L.~Girardello, C.~Kounnas and M.~Porrati, Phys. Lett.
B192 (1987) 368;
\\
I.~Antoniadis, J.~Ellis, E.~Floratos, D.V.~Nanopoulos and T.~Tomaras,
Phys. Lett. B191 (1987) 96;
\\
S.~Ferrara, L.~Girardello, C.~Kounnas and M.~Porrati, Phys. Lett.
B194 (1987) 358.
\item
\label{it}
M.~Dine, N.~Seiberg, X.G.~Wen and E.~Witten, Nucl. Phys. B278 (1986)
769 and B289 (1987) 319;
\\
H.~Itoyama and T.~Taylor, Phys. Lett. B186 (1987) 129;
\\
M.~Cvetic, Phys. Rev. Lett. 59 (1987) 1796
and Phys. Rev. D37 (1988) 2366;
\\
P.~Candelas, X.C.~de la Ossa, P.S.~Green and L.~Parkes, Nucl. Phys. B359
(1991) 21.
\item
\label{ss1}
C.~Kounnas and M.~Porrati, Nucl. Phys. B310 (1988) 355.
\item
\label{ss2}
S.~Ferrara, C.~Kounnas, M.~Porrati and F.~Zwirner,
Nucl. Phys. B318 (1989) 75;
\\
M.~Porrati and F.~Zwirner, Nucl. Phys. B326 (1989) 162.
\item
\label{ss3}
C.~Kounnas and B.~Rostand, Nucl. Phys. B341 (1990) 641;
\item
\label{ss4}
I.~Antoniadis, Phys. Lett. B246 (1990) 377;
\item
\label{ss5}
I.~Antoniadis and C.~Kounnas, Phys. Lett. B261 (1991) 369.
\item
\label{gcond1}
H.-P.~Nilles, Phys. Lett. B115 (1982) 193 and Nucl. Phys. B217 (1983)
366;
\\
S.~Ferrara, L.~Girardello and H.-P.~Nilles, Phys. Lett. B125 (1983)
457.
\item
\label{gcond2}
J.-P.~Derendinger, L.E.~Ib\`a\~nez and H.P.~Nilles, Phys. Lett. B155
(1985) 65;
\\
M.~Dine, R.~Rohm, N.~Seiberg and E.~Witten, Phys. Lett. B156 (1985)
55;
\\
C.~Kounnas and M.~Porrati, Phys. Lett. B191 (1987) 91.
\item
\label{lust}
A.~Font, L.E.~Ib\`a\~nez, D.~L\"ust and F.~Quevedo, Phys. Lett.
B245 (1990) 401;
\\
S.~Ferrara, N.~Magnoli, T.R.~Taylor and G.~Veneziano, Phys.
Lett. B245 (1990) 409;
\\
H.-P.~Nilles and M.~Olechowski, Phys. Lett. B248 (1990) 268;
\\
P.~Bin\'etruy and M.K.~Gaillard, Phys. Lett. B253 (1991) 119;
\\
V.~Kaplunovsky and J.~Louis, preprint UTTG-94-1, LMU-TPPW-94-1 (1994).
\item
\label{astro}
T.R.~Taylor and G.~Veneziano, Phys. Lett. B213 (1988) 450;
\\
J.~Ellis, S.~Kalara, K.A.~Olive and C.~Wetterich, Phys. Lett. B228 (1989)
264;
\\
M.~Gasperini, Torino preprint DFTT-03/94;
\\
M.~Gasperini and G.~Veneziano, preprint CERN-TH.7178/94.
\item
\label{cosmo}
G.D.~Coughlan, W.~Fischler, E.W.~Kolb, S.~Raby and G.G.~Ross,
Phys. Lett. B131 (1983) 59;
\\
J.~Ellis, D.V.~Nanopoulos and M.~Quir\'os, Phys. Lett. B174 (1986)
176;
\\
J.~Ellis, N.C.~Tsamis and M.~Voloshin, Phys. Lett. B194 (1987) 291;
\\
R.~Brustein and P.J.~Steinhardt, Phys. Lett. B302 (1993) 196;
\\
B.~de~Carlos, J.A.~Casas, F.~Quevedo and E.~Roulet, Phys. Lett. B318
(1993) 447;
\\
T.~Banks, D.B.~Kaplan and A.E.~Nelson, Phys. Rev. D49 (1994) 779.
\item
\label{fcnc}
R.~Barbieri and R.~Gatto, Phys. Lett. B110 (1982) 211;
\\
J.~Ellis and D.V.~Nanopoulos, Phys. Lett. B110 (1982) 44;
\\
T.~Banks, Nucl. Phys. B303 (1988) 172;
\\
Y.~Nir and N.~Seiberg, Phys. Lett. B309 (1993) 337.
\item
\label{kpz}
C.~Kounnas, I.~Pavel and F.~Zwirner, preprint CERN-TH.7185/94,
LPTENS-94/08.
\item
\label{effqu}
L.E.~Ib\`a\~nez and H.P.~Nilles, Phys. Lett. B169 (1986) 354;
\\
H.~Itoyama and J.~Leon, Phys. Rev. Lett. 56 (1986) 2352;
\\
E.~Martinec, Phys. Lett. B171 (1986) 2352;
\\
M.~Dine and N.~Seiberg, Phys. Rev. Lett. 57 (1986) 2625;
\\
H.P.~Nilles, Phys. Lett. B180 (1986) 240;
\\
M.~Dine, N.~Seiberg and E.~Witten, Nucl. Phys. B289 (1987) 589;
\\
V.S.~Kaplunovsky, Nucl. Phys. B307 (1988) 145;
\\
L.J.~Dixon, V.S.~Kaplunovsky and J.~Louis, Nucl. Phys. B355 (1991)
649;
\\
J.-P.~Derendinger, S.~Ferrara, C.~Kounnas and F.~Zwirner, Nucl. Phys.
B372 (1992) 145 and Phys. Lett. B271 (1991) 307;
\\
G.~Lopez Cardoso and B.A.~Ovrut, Nucl. Phys. B369 (1992) 351;
\\
I.~Antoniadis, K.S.~Narain and T.~Taylor, Phys. Lett. B267 (1991) 37
and Nucl. Phys. B383 (1992) 93;
\\
I.~Antoniadis, E.~Gava, K.S.~Narain and T.~Taylor, Nucl. Phys. B407
(1993) 706 and Northeastern University preprint NUB-3084.
\item
\label{duality}
K.~Kikkawa and M.~Yamasaki, Phys. Lett. B149 (1984) 357;
\\
N.~Sakai and I.~Senda, Progr. Theor. Phys. 75 (1986) 692;
\\
V.P.~Nair, A.~Shapere, A.~Strominger and F.~Wilczek, Nucl. Phys. B287 (1987)
402;
\\
B.~Sathiapalan, Phys. Rev. Lett. 58 (1987) 1597;
\\
R.~Dijkgraaf, E.~Verlinde and H.~Verlinde, Commun. Math. Phys. 115
(1988) 649;
\\
A.~Giveon, E.~Rabinovici and G.~Veneziano, Nucl. Phys. B322 (1989)
167;
\\
A.~Shapere and F.~Wilczek, Nucl. Phys. B320 (1989) 301;
\\
M.~Dine, P.~Huet and N.~Seiberg, Nucl. Phys. B322 (1989) 301.
\item
\label{flst}
S.~Ferrara, D.~L\"ust, A.~Shapere and S.~Theisen, Phys. Lett. B225
(1989) 363;
\\
S.~Ferrara, D.~L\"ust and S.~Theisen, Phys. Lett. B233 (1989) 147.
\item
\label{ss}
J.~Scherk and J.H.~Schwarz, Phys. Lett. B82 (1979) 60 and Nucl. Phys.
B153 (1979) 61;
\\
E.~Cremmer, J.~Scherk and J.H.~Schwarz, Phys. Lett. B84 (1979) 83.
\item
\label{fayet}
P.~Fayet, Phys. Lett. B159 (1985) 121 and Nucl. Phys. B263 (1986)
649.
\item
\label{rohm}
R.~Rohm, Nucl. Phys. B237 (1984) 553.
\item
\label{decomp}
R.~Rohm and E.~Witten, Ann. Phys. 170 (1986) 454;
\\
M.~Dine and N.~Seiberg, Nucl. Phys. B301 (1988) 357;
\\
T.~Banks and L.J.~Dixon, Nucl. Phys. B307 (1988) 93;
\\
I.~Antoniadis, C.~Bachas, D.~Lewellen and T.N.~Tomaras, Phys. Lett.
B207 (1988) 441.
\item
\label{lrexp}
I.~Antoniadis, C.~Mu\~noz and M.~Quir\'os, Nucl. Phys. B397 (1993) 515;
\\
I.~Antoniadis and K.~Benakli, preprint CPTH-A257-0793 (1993);
\\
I.~Antoniadis, K.~Benakli and M.~Quir\'os, preprint CPTH-A293-0294 (1994).
\item
\label{explo}
V.S.~Kaplunovsky, Phys. Rev. Lett. 55 (1985) 1036;
\\
M.~Dine and N.~Seiberg, Phys. Rev. Lett. 55 (1985) 366;
\\
T.~Taylor and G.~Veneziano, Phys. Lett. B212 (1988) 147.
\item
\label{supext}
S.~Ferrara and B.~Zumino, Phys. Lett. B86 (1979) 279;
\\
S.~Ferrara, C.~Savoy and L.~Girardello, Phys. Lett. B105 (1981) 363.
\item
\label{gaugings}
M.~de Roo, Nucl. Phys. B255 (1985) 515 and Phys. Lett B156 (1985) 331;
\\
E.~Bergshoeff, I.G.~Koh and E.~Sezgin, Phys. Lett. B155 (1985) 71;
\\
M.~de Roo and P.~Wagemans, Nucl. Phys. B262 (1985) 644 and Phys. Lett.
B177 (1986) 352;
\\
P.~Wagemans, Phys. Lett. B206 (1988) 241.
\item
\label{moore}
J.H.~Horne and G.~Moore, Yale preprint YCTP-P2-94.
\item
\label{stdua}
A.~Font, L.E.~Ib\`a\~nez, D.~L\"ust and F.~Quevedo,
Phys. Lett. B249 (1990) 35;
\\
S.J.~Rey, Phys. Rev. D43 (1991) 526;
\\
P.~Binetruy, Phys. Lett. B315 (1993) 80;
\\
R.~d'Auria, S.~Ferrara and M.~Villasante, Class. Quant. Grav.
11 (1994) 481;
\\
J.H.~Schwarz and A.~Sen, Nucl. Phys. B411 (1994) 35 and Phys. Lett.
B312 (1993) 105.
\item
\label{special}
E.~Cremmer, C.~Kounnas, A.~van Proeyen, J.P.~Derendinger, S.~Ferrara,
B.~de Wit and L.~Girardello, Nucl. Phys. B250 (1985) 385;
\\
S.~Cecotti, S.~Ferrara and L.~Girardello, Int. Journ. Mod. Phys. 4 (1989)
2475;
\\
S.~Ferrara and A.~Strominger, in `Strings '89' (R.~Arnowitt et al., eds.),
World Scientific, Singapore, 1989, p.~245;
\\
P.~Candelas and X.C.~de la Ossa, Nucl. Phys. B342 (1990) 246;
\\
A.~Strominger, Commun. Math. Phys. 133 (1990) 163;
\\
L.~Dixon, V.~Kaplunovsky and J.~Louis, as in ref.~[\ref{effcl}];
\\
L.~Castellani, R.~d'Auria and S.~Ferrara, Class. Quant. Grav. 1 (1990) 1767;
\\
S.~Cecotti and C.~Vafa, Nucl. Phys. B367 (1991) 359;
\\
A.~Ceresole, R.~d'Auria, S.~Ferrara, W.~Lerche and J.~Louis, Int. Journ. Mod.
Phys. A8 (1993) 79.
\item
\label{agnt}
I.~Antoniadis, E.~Gava, K.S.~Narain and T.~Taylor,
Northeastern University preprint NUB-3084.
\item
\label{candelas}
P.~Candelas, X.C.~de la Ossa, P.S.~Green and L.~Parkes, as in
ref.~[\ref{it}].
\item
\label{mirror}
B.R.~Greene and M.R.~Plesser, Nucl. Phys. B338 (1990) 15;
\\
P.~Candelas, M..~L\"utken and R.~Schimmrigk, Nucl. Phys.
B341 (1990) 383;
\\
P.~Aspinwall, C.A.~L\"utken and G.G.~Ross, Phys. Lett. B241 (1990)
373;
\\
P.~Aspinwall and C.A.~L\"utken, Nucl. Phys. B353 (1991) 427 and B355 (1991)
482.
\item
\label{blm}
R.~Barbieri, J.~Louis and M.~Moretti, as in ref.~[\ref{softstr}].
\item
\label{rhs}
T.~Banks, D.B.~Kaplan and A.~Nelson, Phys. Rev. D47 (1994) 779;
\\
J.~Bagger, E.~Poppitz and L.~Randall, preprint JHU-TIPAC-940005,
MIT-CTP-2309, NSF-ITP-94-48.
\end{enumerate}
\end{document}